\documentclass[reprint,superscriptaddress]{revtex4-2}

\usepackage{amssymb,amsmath}
\usepackage{graphicx}
\usepackage{multirow}
\usepackage{appendix}
\usepackage{mathtools}
\usepackage{amsmath,bm}
\usepackage{siunitx}
\usepackage{lineno}
\usepackage{physics}
\modulolinenumbers[5]
\graphicspath{{./IMG/}}
\begin{document}

\title [mode = title]{Importance of long-range channel Sr displacements for the narrow emission in Sr[Li$_2$Al$_2$O$_2$N$_2$]:Eu$^{2+}$phosphor}       

\author{Julien Bouquiaux}
\email{julien.bouquiaux@uclouvain.be}
\affiliation{Institute of Condensed Matter and Nanosciences, Universit\'{e} catholique de Louvain, Chemin des \'{e}toiles 8, bte L07.03.01, B-1348 Louvain-la-Neuve, Belgium}
\author{Samuel Ponc\'{e}}
\affiliation{Theory and Simulation of Materials (THEOS), \'Ecole Polytechnique F\'ed\'erale de Lausanne, CH-1015 Lausanne, Switzerland}
\author{Yongchao Jia}
\affiliation{Institute of Condensed Matter and Nanosciences, Universit\'{e} catholique de Louvain, Chemin des \'{e}toiles 8, bte L07.03.01, B-1348 Louvain-la-Neuve, Belgium}
\author{Anna Miglio}
\affiliation{Institute of Condensed Matter and Nanosciences, Universit\'{e} catholique de Louvain, Chemin des \'{e}toiles 8, bte L07.03.01, B-1348 Louvain-la-Neuve, Belgium}
\author{Masayoshi Mikami}
\affiliation{Materials Design Laboratory, Science $\&$ Innovation Center, Mitsubishi Chemical Corporation, 1000,
Kamoshida-cho Aoba-ku, Yokohama, 227-8502, Japan}
\author{Xavier Gonze}
\affiliation{Institute of Condensed Matter and Nanosciences, Universit\'{e} catholique de Louvain, Chemin des \'{e}toiles 8, bte L07.03.01, B-1348 Louvain-la-Neuve, Belgium}
\affiliation{Skolkovo Institute of Science and Technology, Skolkovo Innovation Center, Nobel St. 3, Moscow, 143026, Russia}


\begin{abstract}
The recently discovered Sr[Li$_2$Al$_2$O$_2$N$_2$]:Eu$^{2+}$ red phosphor, candidate for the next generation of eco-efficient white light-emitting diodes, exhibits 
excellent emission spectral position and exceptionally small linewidth. 
It belongs to the UCr$_4$C$_4$-structure family of phosphors, containing many potential candidates commercial phosphors, whose small linewidth, tentatively
ascribed to the high-symmetry cuboid environment of the doping site, has drawn the attention of researchers in the last five years.
We use density functional theory, $\Delta$SCF method 
and configuration coordinate models (CCM) to provide a complete characterization of this material. Using a multi-dimensional CCM, an accurate description of the coupling of the vibronic structure with the electronic 5d$\rightarrow$4f transition is obtained, including the partial Huang-Rhys factors and frequency of the dominant modes.
%
We show that, in addition to the first-coordination shell cuboid deformation mode, low-frequency phonon modes involving chains of strontium atoms along the tetragonal axis shape the emission linewidth in Sr[Li$_2$Al$_2$O$_2$N$_2$]:Eu$^{2+}$. This finding sheds new light on the emission properties of UCr$_4$C$_4$-structure phosphors, possessing similar Ca/Sr/Ba channel.  
Our approach provides a robust theoretical framework to systematically study the emission spectra of such Eu-doped phosphors, and predict candidates with expected similar or even sharper linewidth.
\end{abstract}

\maketitle

\section{Introduction}
\label{sec:Intro}
To mitigate CO$_2$ emission, the use of efficient white-light sources for general lighting is becoming increasingly important worldwide.
In 2019, in its \textit{solid state lighting} plan that relies heavily on light-emitting diodes (LEDs) technology, the US department of energy estimated that the energy savings could account for about a 5\% reduction in the total primary energy budget of the USA~\cite{pattison20202019}.
One way to realize such white light is to coat a primary blue or near ultraviolet LED with one or multiple phosphor material(s) that converts a fraction of the initial high-frequency photons to lower frequency, the resulting light appearing white.
There generally exist two ways to
obtain white light via LEDs with phosphors : (i) a blue LED is used to excite a single yellow phosphor or mixed green and red phosphors;
and (ii) a near ultraviolet LED chip is used to excite the red, green, and blue phosphors \cite{xia2016progress,lin2017inorganic}.

Among the various phosphor compounds, finding a highly efficient red emitting phosphor is crucial for a further increase in the luminous efficacy while keeping a high color-rendering index. 
In order to sufficiently cover the red spectral region while minimizing the efficacy loss caused by the eye poor long-wave sensitivity, red phosphors with narrow emission bandwidth showing a peak in the optimal red region are needed~\cite{pust2015revolution,lin2016critical}. 
In 2019, Hoerder et al.~\cite{hoerder2019sr} reported the experimental discovery of a new narrow-band red-emission phosphor \mbox{Sr[Li$_2$Al$_2$O$_2$N$_2$]:Eu$^{2+}$} (SALON) that fulfills such requirements with an emission peak located at 614~nm, a full width at half maximum (FWHM) of only 48~nm at room temperature and a low thermal quenching of 4\% at operating temperature (420K). 
The fact that the emission peak is located at a shorter wavelength than its cousin Sr[LiAl$_3$N$_4$]:Eu$^{2+}$ (SLA) \cite{pust2014narrow} enables a gain of 16\% in luminous efficacy compared to SLA, while keeping an excellent color rendering and thus enabling a potential leap in the energy efficiency of white-emitting phosphor-converted LEDs. 
SALON belongs to the UCr$_4$C$_4$-structure family of phosphors, from which most interesting narrow-emission candidates have emerged recently~\cite{fang2018control, fang2020cuboid}.

In our work, we first  provide a theoretical characterization of the geometric, electronic and vibrational properties of SALON from first principles, using density-functional theory (DFT).
We study the narrow Eu$_{5d}$$\rightarrow$Eu$_{4f}$ emission spectrum from a theoretical perspective. 
For this purpose, we use the $\Delta$SCF method \cite{chaudhry2011first,chaudhry2014first,jia2016first,jia2017first}, that allows one to compute relevant transition energies and atomic configuration displacement induced by the 5d-4f electronic transition.
In combination with $\Delta$SCF calculations, we use a multi-dimensional configuration-coordinate model with the help of a generating function approach to determine the luminescence intensity spectrum. 
Indeed, this approach allows us to extract the spectral decomposition of the phonon modes participating to the 5d-4f transition by projecting the atomic displacements induced by the 5d-4f transition onto the phonon eigenvectors of the host structure.
For this purpose, we use density-functional perturbation theory (DFPT)~\cite{gonze1997dynamical,baroni2001phonons}  to compute the phonons modes of SALON. 
This approach has been used by Alkauskas \textit{et al.}~\cite{alkauskas2014first} to study the vibronic structure of diamond nitrogen-vacancy (NV) center and, recently, by Linder\"alv \textit{et al.}~\cite{linderalv2020luminescence} to study the lineshape of Ce-doped YAG. 
To our knowledge, the latter work is the only prior existing fully first-principles vibronic
analysis of the 5d-4f emission of lanthanide-based phosphors. We note however that some previous vibronic studies, mixing experimental and theoretical analysis, were already carried out \cite{tanner2003absorption, bachmann2009temperature, kunkel2014bright, lefevre2018electron}.

We compare our result with a one-dimensional configuration-coordinate model (1D CCM).

It was suggested by Hoerder et al. \cite{hoerder2019sr} that the narrow emission band of SALON originates from the Eu$^{2+}$ optical center with high local-symmetry environment composed of a nearly cubic polyhedron of four oxygen and four nitrogen atoms, resulting in an isotropic structural relaxation upon emission. 
This would reduce the number of different energetic states involved in the emission process, leading to a narrow emission band~\cite{hoerder2019sr}. %
Here, we show that this picture is incomplete. 

Our computations indicate that the structural change of the environment of the europium atom when going from excited to ground state is composed of two patterns: 
i) a nearly isotropic expansion of the first coordination shell of the Eu atom composed of O and N, as suggested by Hoerder et al. \cite{hoerder2019sr};
ii) a long-range collective displacement of the Sr channel containing Eu atom away from Eu.
This second pattern couples with low-frequency acoustic and $B_u$ phonon modes in a small energy window, and contributes to the small linewidth. 
It is also responsible for the slow convergence of the results with the supercell size. 
We propose a geometric series model on the Sr displacements to estimate the residual error made.

Overall, we explain the narrow emission spectrum by the combination of i) a small atomic displacement upon emission due to the rigid host structure and ii) an important participation of low-frequency phonon modes in the electronic transition.
We finally find that the shape of our theoretical spectrum agrees well with the experimental one. 

The paper is structured as follows. Section~\ref{sec:Theory} is devoted to theoretical background: we explain how transitions energies and atomic displacements induced by the 5d-4f transition can be used within a configuration-coordinate model framework. 
The generating function approach that allows one to predict luminescence spectra is also introduced. 
In section~\ref{sec:Results}, we present the results of our work. We start with the DFT results on structural and electronic properties of SALON
and present its luminescent properties. We describe which phonon modes dominate the electronic transition. 
We then present the theoretical luminescence spectrum decomposed in all the vibrational modes and we compare it to the spectrum obtained from only one effective vibrational mode. 
Section~\ref{sec:Conclusion} provides additional discussion and concludes this work.
The computational methodology is described in section~\ref{sec:Comp_method}.
%

\section{Theoretical background}
\label{sec:Theory}
\begin{figure}
    \centering
	\includegraphics[width=8.5cm]{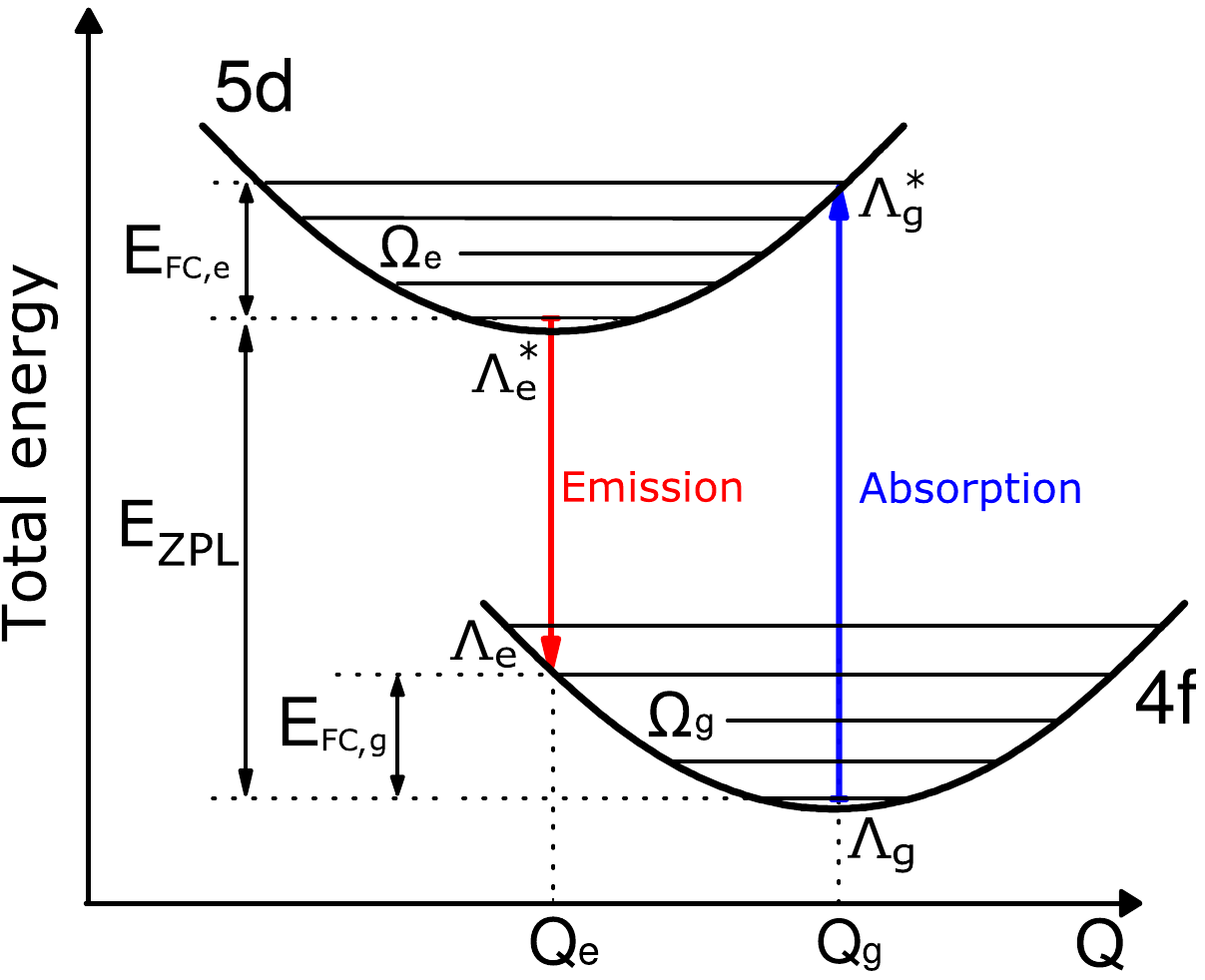}
	\caption{Configuration coordinate diagram, see text for details. }
	\label{fig:1D-CCM}
\end{figure}

The absorption-emission process can be understood based on \textbf{Figure \ref{fig:1D-CCM}}.
We first consider the system in its relaxed 4f ground state ($\Lambda_{\rm g}$) with configuration coordinate $Q_{\rm g}$. 
When a photon is absorbed, the system  is quickly promoted to an unrelaxed excited 5d state ($\Lambda_{\rm g}^*$) without atomic motion. 
Due to this change of electronic state, atomic relaxation towards a new equilibrium configuration $Q_{\rm e}$ takes place with phonon emission. The energy lost in the process is the Franck-Condon shift $E_{\rm FC,e}$. Photon emission occurs then through 5d-4f deexcitation and the system reaches an unrelaxed ground state $\Lambda_{\rm e}$. The cycle is completed by a last relaxation with phonon emission and energy loss $E_{\rm FC,g}$ until the initial state $\Lambda_{\rm g}$.

The total energy loss is the Stokes shift $\Delta S$ given by the sum of the two Franck-Condon shifts $\Delta S=E_{\rm FC,e}+E_{\rm FC,g}$. The zero-phonon line refers to a purely electronic transition with energy $E_{\rm ZPL}$. Ignoring the small difference of zero-point energies between the two states, $E_{\rm ZPL}$ is the difference between the two energy surface minima.

We first describe how to consider the luminescent properties with a 1D-CCM as depicted in Figure \ref{fig:1D-CCM}.
In this model, all the complexity of the nuclei vibrations is reduced to one effective vibrational mode characterized by a generalized configuration coordinate $Q$ that interpolates
linearly between ground- and excited-state atomic coordinates. 
Deviations from linear interpolation 
can be considered when studying
non-radiative decay channels, but have been shown to have a small impact on the luminescence spectrum shape~\cite{Jia-2019-1}.

In the 1D-CCM, the 4f and 5d Born-Oppenheimer potential energies are displayed as a function of the normal coordinate $Q$.
The corresponding ionic configuration is
\begin{equation}
\bm{R}(Q)=
\frac{Q}{\Delta Q}(\bm{R_{\rm e}}-\bm{R_{\rm g}})
+ \bm{R_{\rm g}},
\label{eq:R_Q}
\end{equation}
where $\Delta Q$ is the normal coordinate
change from ground-state to excited-state geometries. 
It writes
\begin{align}\label{eq:Delta_Q_1D}
(\Delta Q)^2 &= \sum_{\alpha i}m_{\alpha}(R_{{\rm e};\alpha i}-R_{{\rm g};\alpha i})^2, 
\end{align}
where $i$ labels Cartesian axes, $\alpha$ atoms, $m_{\alpha}$ atomic masses, $R_{{\rm e};\alpha i}$ and $R_{{\rm g};\alpha i}$ are respectively atomic positions in 5d excited and 4f ground states.
Within such coordinate system, the harmonic ground-state and excited-state energies are respectively
\begin{align}
\label{eq:V_g}
E(Q) &= \frac{1}{2} \Omega_{\rm g}^2 Q^2, \\
E^*(Q) &= \frac{1}{2} \Omega_{\rm e}^2 (Q - \Delta Q)^2 + E_{\rm ZPL}.
\label{eq:V_e}
\end{align}

The associated effective vibrational frequencies in the ground (g) or excited (e) states are
\begin{align}\label{eq:Omega_g_1D}
\Omega_{\rm{\{g,e\}}}^2 &=\frac{2E_{\rm{FC,\{g,e\}}}}{\Delta Q^2}.
\end{align}

We can use Eqs. \eqref{eq:V_g} and \eqref{eq:V_e} to compute the Huang-Rhys factors that indicate the average numbers of phonons emitted during the relaxation process:

\begin{equation}
S_{\rm{\{g,e\}}} = \frac{E_{\rm{FC,\{g,e\}}}}{\hbar\Omega_{\rm{\{g,e\}}}}=\frac{\Omega_{\rm{\{g,e\}}}\Delta Q^2}{2\hbar}.
\label{eq:S}
\end{equation}

Using a semi-classical treatment, one can estimate the FWHM of the spectrum by considering the expectation value of $Q$ and by computing the density of transition as a function of the emitted energy.  At 0~K, the FWHM, W(0), is written as~\cite{henderson2006optical}
\begin{equation}
W(0)=\sqrt{8\ln2}\frac{S_{\rm g}}{\sqrt{S_{\rm e}}}\hbar\Omega_{\rm g}.
\label{eq:W_0}
\end{equation}
This formula is valid for large values of S, such that the otherwise Pekarian  line  shape  of  the emission  spectrum is  well  approximated  by  a  Gaussian  envelope. As will be seen later, this condition is not well fulfilled for SALON phosphor (S $\approx$ 3), but we will use it nonetheless as a reference for later comparison.

Focusing now on the multi-dimensional configuration coordinate model (multi-D CCM), the nuclear motions are a superposition of 3N normal modes of vibration $\nu$ represented by normal coordinates $Q_{\nu}$ and frequency $\omega_{\nu}$ with N the number of atoms in the supercell. 
The harmonic approximation is retained: each phonon mode is independent of each other and the vibrations in the solids are composed of 3N harmonic oscillators. 
The vibrational state $\chi_{\bm{n}}$ is expressed as a product of 3N harmonic oscillator eigenfunctions $\chi_{n_{\nu}}$ with $n_{\nu}$ the vibrational state of the $\nu$-th harmonic oscillator and $\bm{n}$ the set of 3N vibrational state $\{n_1, n_2,...,n_{3N}\}$.  

Making then the Franck-Condon approximation, which states that the transition dipole moment between excited and ground state depends weakly on nuclear coordinates, we write the normalized luminescence intensity at 0~K for a given photon energy $\hbar\omega$ as 
$ L(\hbar\omega)=C\omega^3A(\hbar\omega) $,
with $C$ a normalization constant that is chosen such that the area under the curve is unity, and the emission spectral function is  
\begin{equation}\label{eq:AAA}
A(\hbar\omega)=\sum_{\bm{n}} |\langle \chi_{{\rm g},\bm{
n}}|{\chi_{{\rm e},\bm{0}}}\rangle|^2\delta(E_{\rm ZPL}+ E_{{\rm g},\bm{n}}-\hbar\omega),
\end{equation}
with $E_{{\rm g},\bm{n}}=\sum_{\nu}n_{\nu}\hbar\omega_{{\rm g},\nu}$, the energy of the state $\chi_{{\rm g},\bm{n}}$ \cite{alkauskas2014first}.
Note that, in the 1D-CCM and assuming the same harmonic curvature in the ground and excited states, $\Omega_{\rm g}=\Omega_{\rm e}$, such that $S_{\rm{g}}=S_{\rm{e}}=S$, Equation~\eqref{eq:AAA} reduces to 
\begin{equation}\label{eq:AAA_1D}
A(\hbar\omega)=\sum_n e^{-S}\frac{S^n}{n!} \delta(E_{\rm ZPL}+ n\hbar\Omega-\hbar\omega).
\end{equation}

In the multi-D CCM approach, the harmonic approximation allows one to write the 3N-dimensional Franck-Condon overlap as a product of 3N overlap integrals $\langle \chi_{{\rm g},\bm{n}}|\chi_{{\rm e},\bm{m}}\rangle = \prod_{\nu}\langle \chi_{{\rm g},n_\nu}|\chi_{{\rm e},m_\nu}\rangle$. 
Indeed at 0~K, all initial vibrational states are in their lowest state $\bm{n}=\bm{0}$. 
We finally suppose that the Born-Oppenheimer curvatures are the same in the $4f$ and $5d$ state which holds for SALON.
%
This allows one to further simplify Equation~\eqref{eq:AAA} since $|\langle \chi_{{\rm g},n_{\nu}}|\chi_{{\rm e},0}\rangle|^2 = e^{-S_\nu}\frac{S_{\nu}^{n_{\nu}}}{n_{\nu}!}$ with $S_{\nu}$ the partial Huang-Rhys factor of mode $\nu$~\cite{henderson2006optical}. 
We then write the transition matrix element as 
\begin{equation}\label{eq:matrix_el}
|\langle\chi_{4f,\bm{n}}|\chi_{5d,\bm{0}}\rangle|^2=\prod_{\nu=1}^{3N}e^{-S_\nu}\frac{S_{\nu}^{n_{\nu}}}{n_{\nu}!}.
\end{equation}

Quantities that enters Eqs.~\eqref{eq:AAA} and \eqref{eq:matrix_el} are phonon frequencies $\omega_{\nu}$ and partial Huang-Rhys factors $S_{\nu}$. 
Following the work of Alkauskas \textit{et al.}~\cite{alkauskas2014first},
we define the weight by which each mode $\nu$ contributes to the atomic position changes when a electronic transition occurs as $p_{\nu}=(\Delta Q_{\nu}/\Delta Q)^2$, where 
\begin{align}\label{eq:Delta Q_multiD}
\Delta Q_{\nu} &= \sum_{\alpha,i} m_{\alpha}^{1/2}(R_{{\rm e};\alpha i}-R_{{\rm g};\alpha i})\Delta r_{{\nu};\alpha i}, \\
(\Delta Q)^2   &= \sum_{\nu} (\Delta Q_{\nu})^2,
\label{eq:DeltaQ}
\end{align}
and $\Delta r_{{\nu};\alpha i}$ is the vector that represents the displacement of atom $\alpha$ in phonon mode ${\nu}$. 
This vector is normalized such that $\sum_{\alpha,i} \Delta r_{{\nu};\alpha i} \Delta r_{l;\alpha i}=\delta_{{\nu},l}$. 
In other words, the contribution of a mode to the electronic transition is given by the projection of the phonon eigenvector associated with this mode on the atomic distortion induced by this electronic transition, weighted by the atom mass. 
It is possible to define an effective frequency as
\begin{equation}\label{eq:Omega_multiD}
\Omega_{\rm eff}^2=\sum_{\nu} p_{\nu} \omega_{\nu}^2,
\end{equation}
where $\omega_{\nu}$ is the frequency of the mode ${\nu}$. 
The partial Huang-Rhys factor associated with the mode ${\nu}$ indicates the number of phonons of mode $\nu$ involved in the 4f-5d transition. 
Assuming harmonicity, we have 
\begin{equation}\label{eq:S_nu}
S_{\nu}=\frac{\frac{1}{2}\omega_{\nu}^2\Delta Q_{\nu}^2}{\hbar\omega_{\nu}}=\frac{\omega_{\nu} \Delta Q_{\nu}^2}{2\hbar}.
\end{equation}
By introducing a Huang-Rhys spectral decomposition~\cite{miyakawa1970phonon},
\begin{equation}
S(\hbar\omega)=\sum_{\nu} S_{\nu} \delta(\hbar\omega - \hbar\omega_{\nu}),
\label{eq:S(hw)}
\end{equation}
a generating function approach yields the spectral function $A(\hbar\omega)$
\begin{equation}
A(E_{\rm ZPL}-\hbar\omega) = \frac{1}{2\pi\hbar}\int_{-\infty}^{+\infty} G(t)e^{i\omega t-\gamma\abs{t}}dt,
\label{eq:A(hw)_generating}
\end{equation}
where the generating function $G(t)$ is 
\begin{align}
G(t) &= e^{S(t)-S(0)}, \\
S(t) &= \int_0^{+\infty}S(\hbar\omega)e^{-i\omega t} d(\hbar\omega).
\end{align}
The $S(0)=\sum_{\nu}S_{\nu}$ is the total Huang-Rhys factor and the parameter $\gamma$ represents the homogeneous Lorentzian broadening of each transition. 
One can also model inhomogeneous broadening due to ensemble averaging with a convolution of $A(\hbar\omega$) with a Gaussian. 
Within this multi-dimensional approach, calculating the emission spectrum requires to compute in DFT: (i) the zero-phonon line energy $E_{\rm ZPL}$, (ii) the atomic configuration change induced by 4f-5d transition $(R_{{\rm e};\alpha,i}-R_{{\rm g};\alpha,i})$, (iii) the phonon eigenvectors $\Delta r_{{\nu};\alpha i}$ and eigenfrequencies $\omega_{\nu}$. 
The first two are accessible through the $\Delta$SCF method while DFPT can deliver the third one.
Within the 1D-CCM, computing all phonon modes with DFPT is not needed, only one relevant effective phonon frequency is deduced from the harmonic approximation, the knowledge of the Franck-Condon shift and the normal coordinate change $\Delta Q$.
%
\section{Results and discussion}
\label{sec:Results}

\subsection{Structural properties}
\label{subsec:Structural}


Based on single-crystal X-ray diffraction data, Hoerder \textit{et al.}~\cite{hoerder2019sr} showed that SALON crystallizes in a tetragonal phase (space group $P4_2/m$) with the unit-cell parameter $a=b=7.959$~\si{\angstrom} and $c=3.184$~\si{\angstrom}  with 2 formula units per cell and 18 atoms per primitive cell. 
\begin{figure}
\centering
\includegraphics[width=8.5cm]{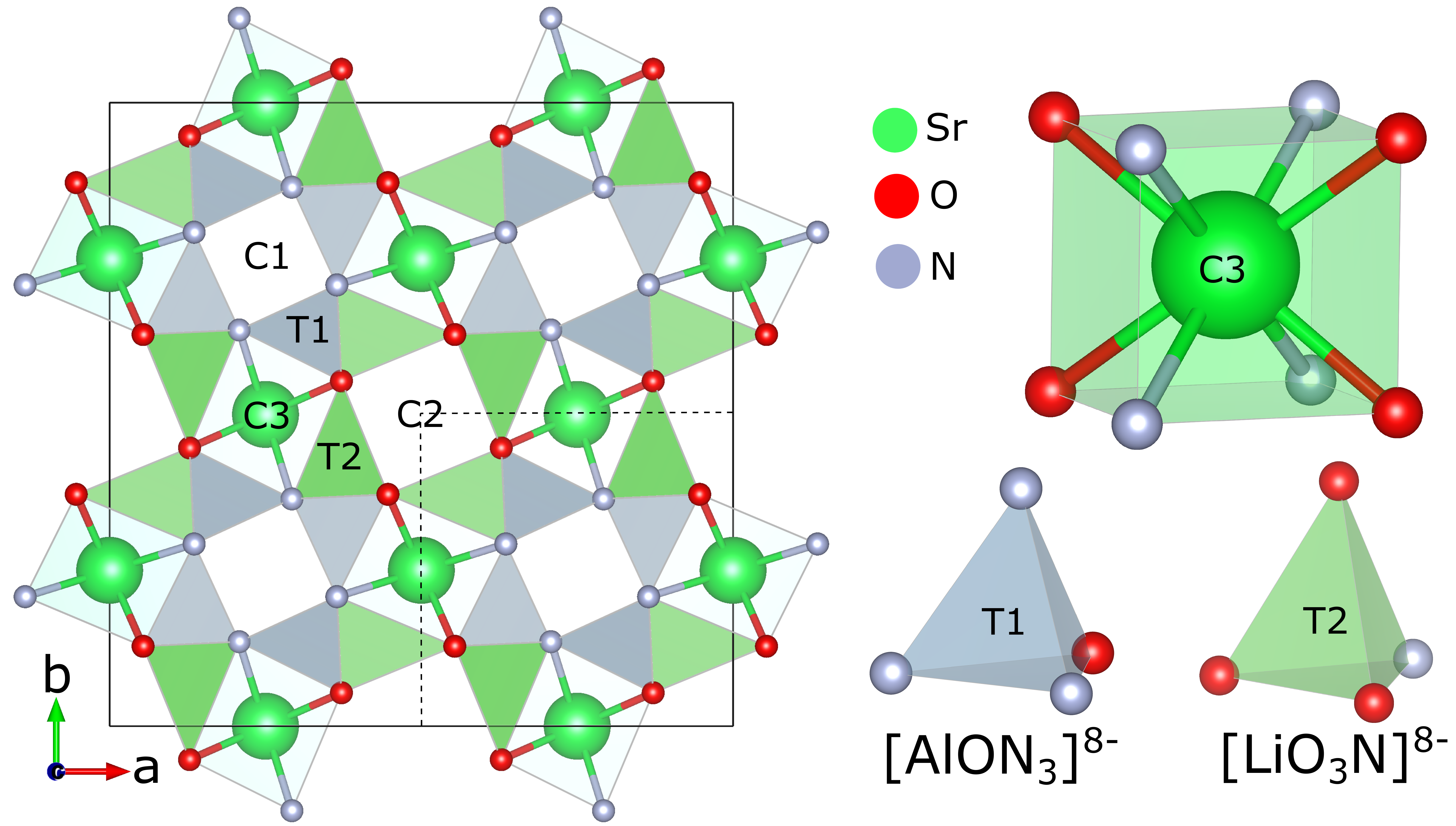}
\caption{2x2x2 supercell of undoped SALON.  Tetrahedra T1 are composed of aluminium coordinated with three nitrogen atoms and one oxygen atom forming a [AlON$_3$]$^{8-}$ unit. Tetrahedra T2 are composed of lithium coordinated with one nitrogen atom and three oxygen atoms forming a [LiO$_3$N]$^{8-}$ units. These tetrahedra compose a network of vierer rings arranged in three types of channels of which two are empty (C1 and C2) and the third hosts strontium cations that are coordinated with four oxygen atoms and four nitrogen atoms in a nearly cubic environment. The unit cell is represented with dashed lines.}
\label{fig:undoped_crystal_structure_222}
\end{figure}
The structure is a variant of the UCr$_4$C$_4$ structure type, well known in the field of phosphor materials to provide a highly condensed and rigid framework structure~\cite{zhao2018next,liao2018learning,fang2020cuboid}, with strontium on the corresponding uranium site, aluminium and lithium on the corresponding chromium site, and nitrogen and oxygen on the corresponding carbide site.
Two tetrahedra types form a network hosting the strontium cations in one of the resulting channels shown in \textbf{Figure~\ref{fig:undoped_crystal_structure_222}}. 
%
%
%
%
Strontium atoms are coordinated with four nitrogen and four oxygen atoms which gives a highly-symmetrical cube-like coordination.

The structural parameters of undoped SALON computed within DFT are listed in \textbf{Table~\ref{tab:Structural_parameters}}. 
We observe globally a slight overestimation of the structural parameters by about 0.5-0.7~\% which is common of GGA-PBE functionals. Since our computation neglects all vibrational and anharmonic effects, while the experimental structure is measured at room-temperature, this error might be underestimated because of zero-point and thermal expansion effects.
When a Sr atom is replaced by an Eu atom in the doped configuration, with decrease of symmetry to point group $2/m$ ($C_{2h}$), we observe that the Eu-O and Eu-N bond lengths are quite close to the Sr-O and Sr-N bond lengths, which is reasonable since the experimental atomic radii of Eu and Sr in their VIII coordination state are very close; 1.39 and 1.40~\si{\angstrom}, respectively~\cite{shannon1976revised}. In contrast, we also observe that the Eu-Sr length is more than 2 $\%$ bigger than the Sr-Sr length in the undoped structure.

\begin{table*}
\centering
\begin{tabular}{llllll}
\hline \hline
 & Experiment~\cite{hoerder2019sr} & \multicolumn{3}{c}{This work} \\ 
\cline{2-5} 
                  & Undoped & Undoped & Doped 4f &  Doped 5d  \\ 
\cline{2-5} 
a [\si{\angstrom}]   & 7.959   & 8.013 (+0.67\%)  & 8.012 (-0.01\%)    & 8.012 (+0\%) \\
b [\si{\angstrom}]   & 7.959   & 8.013 (+0.67\%)  & 8.012 (-0.01\%)    & 8.012  (+0\%)\\
c [\si{\angstrom}]   & 38.208   & 38.438 (+0.60\%)  & 38.442 (+0.01\%)    & 38.442 (+0\%)     \\
X-Sr length [\si{\angstrom}] & 3.184 & 3.131 (-1.66\%)   & 3.206 (+2.39\%)   & 3.171 (-1.10\%)  \\
X-O length [\si{\angstrom}] & 2.659 & 2.667 (+0.30\%)  & 2.678 (+0.41\%)   & 2.633 (-1.71\%)  \\
X-N length [\si{\angstrom}] & 2.760 & 2.782 (+0.79\%)  & 2.784 (+0.07\%)  & 2.738 (-1.68\%)    \\
O-X-O angle [$^{\circ}$] & 73.572 & 73.819 (+0.34\%) & 73.719 (-0.13\%)   & 72.514 (-1.66\%)  \\
N-X-N angle [$^{\circ}$] & 70.446  & 70.301 (-0.21\%)  & 70.356 (+0.08\%) & 70.628 (+0.39\%) \\
Polyhedral volume [\si{\angstrom}$^3$]  & 30.411    & 30.870 (+1.49\%)     & 31.107 (+0.76\%)                 & 29.715 (-4.68\%)  \\ 
\hline \hline
\end{tabular}
\caption{Structural parameters of undoped and doped SALON, either in ground 4f or excited 5d state for a 1$\times$1$\times$12 supercell. The X symbol refers to strontium when undoped and europium when doped. Polyhedral volume refers to the nearly cubic environment of Sr/Eu composed of 4 oxygen and 4 nitrogen. The percentages in parentheses are calculated with respect to the previous column.}
\label{tab:Structural_parameters}
\end{table*}

\begin{figure*}
\centering
\includegraphics[width=17.8cm]{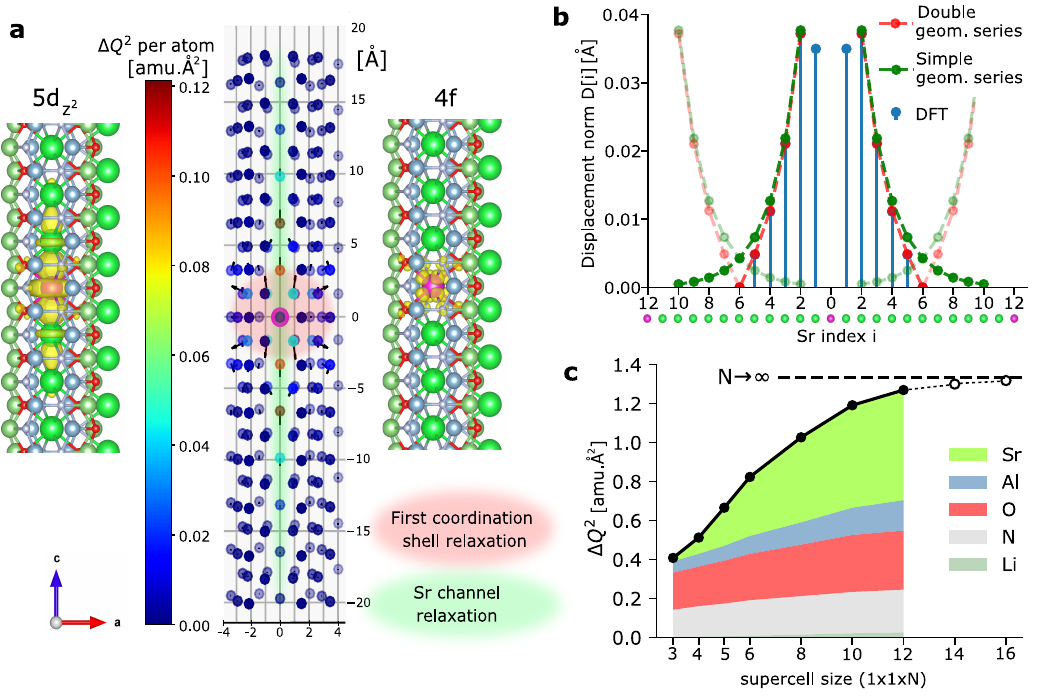}
\caption{a) Excited state geometry and its highest 5d$_{z^2}$ occupied orbital (left) compared to the ground state geometry and its highest 4f occupied orbital (right). The isosurface levels are the same.
The atomic displacements induced by the 5d to 4f electron density change is displayed in the middle of the figure. Central pink sphere represents Eu atom.
The vector's magnitude is multiplied by 25 for visualization.
The color associated to each atom provides its contribution to the total normal coordinate change $(\Delta Q)^2$.
Shaded areas indicates the two identified relaxation patterns. b) Blue points are DFT displacements induced by the 5d-4f transition of the i-th distant Sr from Eu, projected on c-axis, in a 1$\times$1$\times$12 supercell.
The red curve is a sum of two counter-acting geometric series that models the effect of periodically arranged Eu. The green curve represents the displacements if Eu atom was completely isolated. Transparent curves show the displacements associated to the first periodic replica. c) Convergence of $(\Delta Q)^2$ with the supercell size 1$\times$1$\times$N with an atomic decomposition. Dashed black line represents the estimated converged value when N$\rightarrow$ $\infty$. Empty circles estimate the values for larger supercells}
\label{fig:4f-5d_relaxation}
\end{figure*}
To understand the relaxation process accompanying the change in electronic density induced by the 5d-4f transition, a comparison is performed between the $4f$-state geometry ($\Lambda_{\rm g}$, $\Lambda_{\rm g}^{*}$)  and the $5d$-state geometry ($\Lambda_{\rm e}^{*}$, $\Lambda_{\rm e}$), see \textbf{Figure~\ref{fig:4f-5d_relaxation}}. %
On the left and right of Figure~\ref{fig:4f-5d_relaxation}a, we present the highest occupied Kohn-Sham orbitals (norm of the wavefunctions) associated to the 4f$^7$($^8$S$_{7/2}$) ground and 4f$^6$($^7$F$_J$)5d$^1$ excited states.
We can directly notice how the 5d orbital is much more spatially extended than the 4f orbital. 
On the middle of  Figure~\ref{fig:4f-5d_relaxation}a, we present the displacement field associated to the 5d-4f transition. 
The black vectors indicate the atomic displacements when going from excited to ground state $(\bm{R}_{{ e},\alpha}-\bm{R}_{{ g},\alpha})$ while the color associated to each atom provides its contribution to the total normal coordinate change $(\Delta Q_{\alpha})^2 = \sum_{i}m_{\alpha}(R_{{\rm e};\alpha i}-R_{{\rm g};\alpha i})^2$.
A more quantitative comparison between the ground and the excited state environment is provided in Table~\ref{tab:Structural_parameters}. 

From these results, we identify two important relaxation patterns when going from 5d to 4f state. 
First, we observe a nearly isotropic expansion of the first coordination shell of Eu composed of four N and four O arranged in a nearly cubic environment (red-shaded area of Figure \ref{fig:4f-5d_relaxation}a). See the comparison between Eu-N length, Eu-O length, and polyhedral volume of excited and ground states. We justify the isotropic character of this expansion with the small O-Eu-O and N-Eu-N angle changes: no significant distortion effect is observed. 
In reference~\cite{joos2020insights}, the following explanation is given to explain this expansion effect: the bond length between an f-element ion in a $4f^N$ configuration and ligands is realized by the interaction between the $5p^6$ shell and the ligand's valence electrons. 
According to this reference, when the ion is in a $4f^{N-1}5d^{1}$ configuration, additional covalent interactions appear from the ligand to the inner 4$f$ hole that shortens the bond length. 
We could also argue that the excited $Eu^{2+}$ in the $4f^{6}5d^{1}$ configuration gets closer to $Eu^{3+}$ configuration, leading to stronger Eu-O and Eu-N bonds, due to both Coulomb and covalent strengthening.
This uncommon situation of a bond length shortening upon $f \rightarrow d$ excitation has been already observed experimentally and studied theoretically in various other lanthanide compounds \cite{meijerink1989luminescence,barandiaran2003quantum,tanner2003absorption, mahlik2009pressure,mahlik2009temperature, de2016new}.

Second, we observe a long-range relaxation of the strontium atoms belonging to the strontium channel containing the europium atom (see green shaded area of Figure \ref{fig:4f-5d_relaxation}a). When going from 5d to 4f state, those strontium atoms are pushed away from europium along c-axis.
Comparing first the displacement magnitude of the two patterns, we observe that the displacements of the first distant Sr is slightly smaller than the O/N displacement associated to the first pattern (see supplementary \textbf{Figure S10} for a visualisation of the displacement norm as a function of the distance from europium). 
However, what matters in the luminescent properties is the normal coordinate change $(\Delta Q)^2$ (i.e. the displacement weighted by the atom masses), which explains why this second pattern is actually the dominant one, the Sr being heavier. 
It appears that about half of the total normal coordinate change comes from Sr displacements, see Figure \ref{fig:4f-5d_relaxation}c.

It is surprising to observe that the displacement of the first-distant Sr is slightly smaller than the second-distant Sr. We attribute this effect to the large spatial extend of the 5d$_{z^2}$ orbital. 
Beyond the second distant Sr, the Sr displacements follow an exponential decay, as explained later.

We performed a convergence study of $(\Delta Q)^2$ as a function of the supercell size (see Figure \ref{fig:4f-5d_relaxation}c and supplementary \textbf{Figure S1}). Increasing the number of atoms in the x-y plane ($\sqrt{2}\times\sqrt{2} \times $ N supercells) does not affect the result while increasing the Eu-Eu distance along c-axis does. 
Indeed, for too small $1\times1\times$ N supercells, there is a spurious cancellation of Sr displacements caused by periodically repeated Eu. The c-component of the supercell size (or equivalently the Eu-Eu distance) required to obtain a strictly converged $(\Delta Q)^2$ is very large, and instead,
we model this cancellation effect by expressing the Sr displacements $D[\rm{i}]$ in [\si{\angstrom}] unit as a sum of two counter-acting decreasing geometric series,
\begin{equation}
    D[\rm{i}]=C(A^{(\rm{i}-2)}-A^{(N_{Sr}-1-\rm{i})}),
    \label{eq:Sr_dis}
\end{equation}
where i labels the Sr atoms, $N_{Sr}$ denotes the number of Sr atoms between two adjacent Eu, which is 11 in the case of the 1$\times$1$\times$12 supercell used, and where C and A are the fitting coefficients.
With A = 0.58 [/] and C = 0.037 [\si{\angstrom}], the DFT displacements are well reproduced with this simple model, see red curve of Figure \ref{fig:4f-5d_relaxation}b. On this figure, plain and transparent green curves refers respectively to the first and second terms of Equation \ref{eq:Sr_dis}. 
Note that we excluded the first Sr because of the above-mentioned reasons.
If we now consider a perfectly isolated Eu, the Sr displacements would be expressed as a simple decreasing geometric series with the same fitting coefficients A and C (plain green curve). 
We compute the error made as the difference between the red and green displacements. 
In term of normal coordinate change, this gives a difference of 0.028 [amu.$\si{\angstrom}^2$], a 4.8 $\%$ error with respect to the Sr contribution of the total $(\Delta Q)^2$ . 
Indeed, these additional Sr displacements leads to small additional displacements of their atomic neighbours that we simply model with this same percentage error. 
In total, we estimate the strictly converged value (i.e. without finite supercell size effect) to be $(\Delta Q)^2 \approx 1.33$ [amu.$\si{\angstrom}^2$], 4.7 \% higher than the computed DFT value. 
The  impact on the luminescent properties of a specific channel structure along one crystallographic direction  is a rarely observed effect. We still note some works on the luminescence of specific compounds with this 1D feature \cite{ramanantoanina2015prospecting, schoneborn2007crystal}. 
To our knowledge however, it is the first time that such long-range channel relaxation is theoretically highlighted and analysed in phosphor materials. As it will be seen in sub-section \ref{subsec:Lum}, this greatly impacts the computed luminescent properties and it might be a common denominator that explains the particular luminescent properties of this type of phosphor materials. 

\subsection{Electronic properties}
\label{subsec:Elec}


\begin{figure}
\centering
\includegraphics[width=8.5cm]{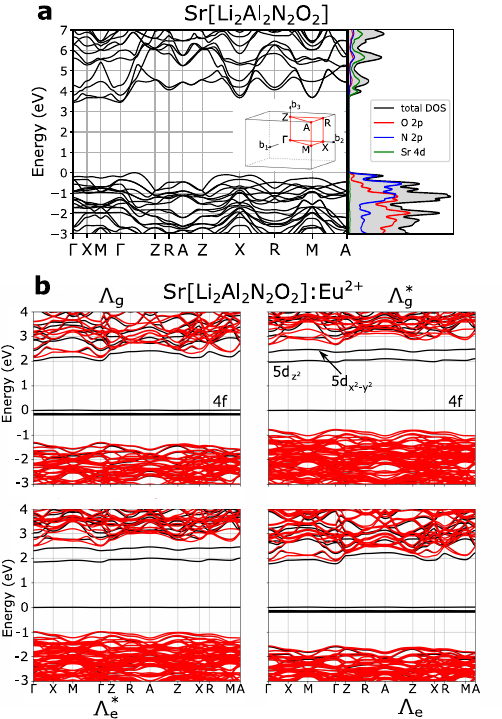}
\caption{a) Electronic band structure of undoped SALON near the fermi level and the associated partial density of states. b) Electronic band structure of SALON:Eu. $\Lambda_{\rm g}$, $\Lambda_{\rm g}^*$, $\Lambda_{\rm e}^*$, $\Lambda_{\rm e}$ refers to the notation used in Figure \ref{fig:1D-CCM}. 
Black and red curves distinguishes  spin up from spin down electrons, respectively.}
\label{fig:electronic_structure}
\end{figure}
The electronic band structure and the partial density-of-states near the Fermi level of undoped SALON are presented in \textbf{Figure~\ref{fig:electronic_structure}a}, obtained with the relaxed cell parameters. 
For the electronic structure over the entire energy range, see supplementary \textbf{Figure~S11}.

The computed indirect band gap of undoped SALON is 3.44~eV, underestimating the experimental gap of 4.4 eV~\cite{hoerder2019sr}, a quite standard
underestimation, often observed with local or \mbox{(semi-)local} exchange-correlation functionals, like the presently used GGA-PBE.
Figure~\ref{fig:electronic_structure}b shows the band structures of the four characteristic states of the one-dimensional coordinate model (see Figure~\ref{fig:1D-CCM}) for a 1$\times$1$\times$5 supercell size.  We find that in contrast to the emission spectrum, the electronic band structure converges faster with the supercell size.
In the ground state, the seven $4f$ states are correctly placed inside the band gap.
In the excited state ($\Lambda_{\rm g}^*$ or $\Lambda_{\rm e}^*$), only one $4f$ state is left in the band gap. This change is a typical feature of the DFT+U approach. 
Then, we observe one first $5d$ state, presenting mainly a $5d_{z^2}$ character,
actually a superposition of $l=2$,
$m=-2,0$ and $2$ angular momentum components, belonging to the 
a$_g$ irreducible representation 
of the above-mentioned 2/m point group.
It shows a small dispersion along the c-axis coming from a residual interaction between periodically repeated europium atoms. 
See supplementary \textbf{Figure S2} for the dispersion evolution with the supercell size. 
The next higher band lying in the band gap is identified as mainly $5d_{x^2-y^2}$. 
When in the excited state, the $4f$-like hole state can also be analyzed in terms of its angular momentum components, showing a superposition of $l=3$, $m=-3, -1, 1$ and $3$ components, so belonging to the 
b$_u$ irreducible representation 
of the above-mentioned 2/m point group.

\subsection{Luminescent properties}
\label{subsec:Lum}


\begin{table*}
\centering
	\begin{tabular}{l|lllll}
		\hline \hline
		&$E_{\rm ZPL}$ &  $E_{\rm FC,g}$    & $E_{\rm FC,e}$      & $\Delta S$ \\ 
		&(eV) &(eV)&(eV)&(eV) \\ \hline 
		SALON:Eu&  2.032  & 0.096     &    0.086        &    0.182       \\ 
		SLA:Eu 1   &2.038   &      0.076       &      0.057       &   0.133     \\  
		SLA:Eu 2&  2.049&      0.060           &    0.111         &     0.171       \\ \hline \hline
		&$\Delta Q$      & $\hbar\Omega_{\rm g}$ & $\hbar\Omega_{{\rm e}}$ & $S_{\rm g}$   & $S_{\rm e}$   \\ 
		& ($\sqrt{amu} \si{\angstrom}$) &(meV) / (cm$^{-1}$)&(meV) / (cm$^{-1}$)&/&/ \\
		\hline 
		SALON:Eu&   1.127 &   25.15 / 202.87      &23.86 / 192.48           &  3.82         &    3.62     \\ 
		SLA:Eu 1&0.756&   33.40 / 269.39           &     29.0 / 233.90        & 2.63        &   1.71       \\ 
		SLA:Eu 2 &  1.222  &  18.30 / 147.60       &  25.0 / 201.63  & 3.27& 4.46 \\ \hline \hline
	\end{tabular}
	\caption{DFT parameters used in the 1D-CCM model. A comparison is provided with the Sr[LiAl$_3$N$_4$]:Eu$^{2+}$ (SLA) compound (two non-equivalent Sr sites, see Ref.~\cite{jia2017first})}
	\label{tab:1D_CCM_results}
\end{table*}
We compute the luminescence properties of doped SALON and report the DFT parameters required for the 1D model in \textbf{Table~\ref{tab:1D_CCM_results}}. 
A comparison with the similar compound Sr[LiAl$_3$N$_4$]:Eu$^{2+}$ (SLA) is provided using data from reference~\cite{jia2017first}, where a similar computational methodology was used.
The experimental $E_{\rm{ZPL}}$ is difficult to estimate precisely as no vibronic structure is present in the 15K-spectrum \cite{hoerder2019sr}. 
We place it in the range 2.03 - 2.08 eV.
Our computed zero-phonon line energy $E_{\rm{ZPL}}=2.032$ eV falls within this range. 
The convergence of $E_{\rm{ZPL}}$ as a function of the supercell size can be found in supplementary \textbf{Figure S3} and shows that 0.1 \% convergence is achieved with respect to the 1$\times$1$\times$8 supercell.

Then, comparing with the 15 phosphors studied in reference~\cite{jia2017first}, we notice that SALON exhibits a small Stokes shift $\Delta S$ and and a small $\Delta Q$. 
This indicates that the geometry of the system is not much affected by the 5d-4f transition. 
We observe also that the effective vibrational energies in the ground (25.15 meV) and excited state (23.86 meV) are close to each other, meaning that the curvature of the Born-Oppenheimer potential in the ground and excited state are quite similar, in contrast to SLA (see table \ref{tab:1D_CCM_results}).

We also note that the Huang-Rhys factor and phonon frequency  are consistent with the Stokes shift. As explained in reference~\cite{de2015resolving}, $\Delta S=0.182$ eV should lie between $(2S_{\rm g}-1)\hbar\Omega_{{\rm g}} = 0.167$~eV and $2S_{\rm g}\hbar\Omega_{{\rm g}} = 0.192$~eV, which is the case.
Finally, the semi-classical FWHM = 119 meV obtained from Equation~\eqref{eq:W_0} is quite close to the experimental FWHM = 135 meV at low temperature.

\begin{figure*}
	\centering
	\includegraphics[width=17.8cm]{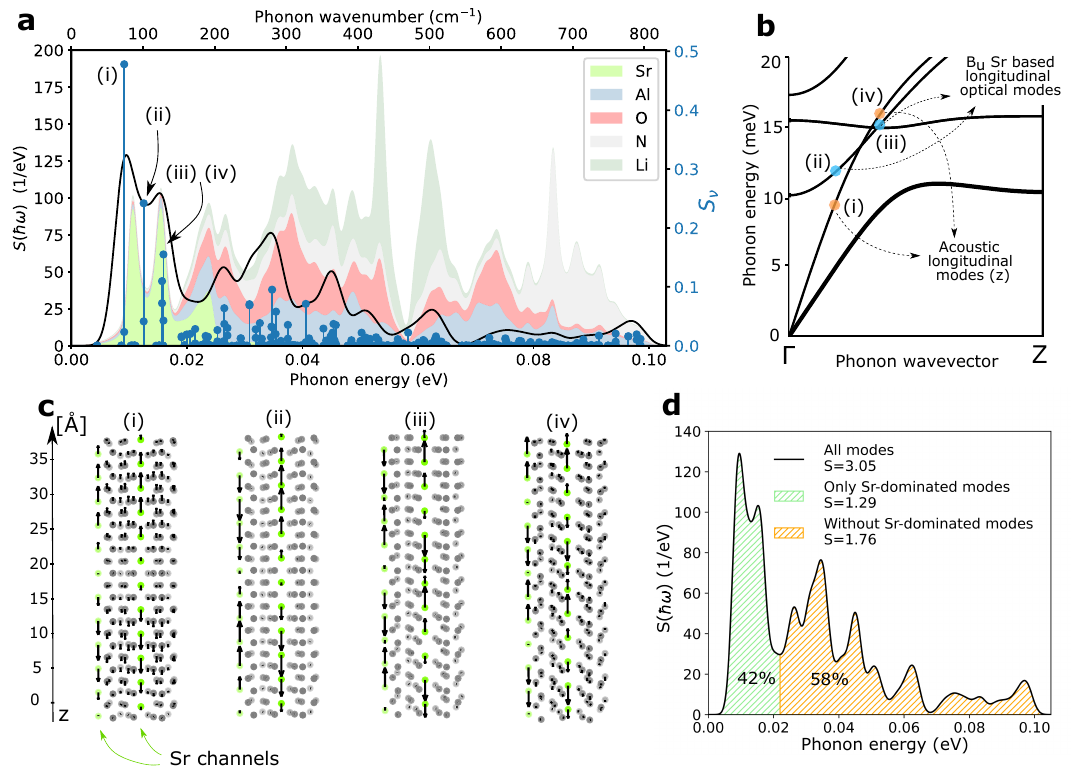}
	\caption{a) Partial Huang-Rhys factors $S_{\nu}$ (Equation~\eqref{eq:S_nu}) and spectral decomposition $S(\hbar\omega)$ (Equation~\eqref{eq:S(hw)}) using a 4~meV Gaussian broadening. The background shows the partial phonon density-of-states of undoped SALON. b) Phonon dispersion at low phonon energy between $\Gamma$-$Z$. High-coupling modes are highlighted. c) Visualisation of the eigenvectors corresponding to the low-frequency high-coupling modes. d) Effect of Sr-dominated modes on the spectral decomposition.}
	\label{fig:S_(hw)}
\end{figure*}

We now present the results from the multi-dimensional approach described in Section~\ref{sec:Theory}.
The discrete partial Huang-Rhys factors $S_{\nu}$ are shown in \textbf{Figure~\ref{fig:S_(hw)}a} along with the spectral decomposition $S(\hbar\omega)$ (Gaussian smearing of 4 meV). 
The phonon density of states is displayed in the background of this figure. 
See supplementary informations for related computed properties (phonon band structure, phonon frequencies at $\Gamma$, Born effective charges and dielectric tensor). 

This reveals the nature of the electron-phonon coupling associated to the 5d-4f transition : At low  energy ($<$ 0.022~eV), phonon modes associated to Sr atoms dominates. This threshold is fixed such that, at 0.022~eV, Sr atoms contribute to half of the phonon density of states. It is also where a dip is observed in the spectral function. We highlight in this low-energy region the four modes presenting the highest partial Huang-Rhys factors $S_{\nu}$. We present them in the phonon dispersion curve along $\Gamma$-$Z$ in Figure \ref{fig:S_(hw)}b and we show their eigendisplacements in Figure \ref{fig:S_(hw)}c where the displacements have been scaled by a factor 10 for clarity.  
The mode that couples the most (i) is an acoustic mode directed along z, of high wavelength. 
On this same phonon branch, the acoustic mode (iv) with half-wavelength also couples strongly. 
Mode (ii) presents a $B_u$ representation where strontium atoms of different channels are out of phase. 
Mode (iii) corresponds to the same mode but with half-wavelength. 
Indeed, those long-wavelength collective displacements of the Sr atoms along the c-axis couple strongly with the above-described 5d-4f relaxation of the Sr channel containing Eu.

Above 0.022 eV, the spectral decomposition is made of a multitude of modes containing Al, O and N displacements. We still notice the large peak around 0.035 eV composed of modes that couple strongly with the relaxation of the first coordination shell of Eu (expansion of the nearly cubic environment of Eu). 
This coupling to a large number of modes instead of one single effective ``breathing" mode coincides with the results of Linderälv \textit{et al.} \cite{linderalv2020luminescence} where a similar feature was found with YAG:Ce$^{3+}$ phosphor. In Figure \ref{fig:S_(hw)}d, we quantify the importance of Sr-dominated phonon modes by computing their contribution to the total Huang-Rhys factor and found out that the Sr-dominated modes contributes 42\% to the narrow emission in  Sr[Li$_2$Al$_2$O$_2$N$_2$]:Eu$^{2+}$.

For completeness, the convergence of this spectral decomposition as a function of the supercell size is presented in supplementary \textbf{Figure S4} and \textbf{S5}. 
It indicates that the result is well converged for high phonon energies while the exact shape of the low energy part of $S(\hbar\omega)$, composed of modes involving Sr atoms, is not yet totally converged. 
By increasing the supercell size further, one would simply displace the points highlighted in Figure \ref{fig:S_(hw)}b towards $\Gamma$ on their corresponding branches. 
This concern reflects first the difficulty to obtain a strict convergence of the structural relaxation of the Sr channel, as described in section \ref{subsec:Structural}. Second, due to the coupling to high-wavelength acoustic phonon modes, obtaining a strict convergence of the low-energy shape requires also to obtain a denser sampling of those phonons. Particular embedded methods to artificially increase the supercell size can be used to tackle this problem \cite{alkauskas2014first,razinkovas2020vibrational}. We reserve such improvements for future work.
However, we note that the integrated spectral function (i.e. $S=\sum_{\nu}S_{\nu}$) presents a quicker convergence. In addition, this lack of strict convergence at low-energy has only a small impact on the smeared emission intensity $L(\hbar\omega)$ due to the counterbalancing effect of a decreasing total Huang-Rhys factor and an increasing effective frequency (see Figure S5)

%

%

The computed effective frequency defined in Equation~\eqref{eq:Omega_multiD}, 25.5~meV, agrees well with the effective frequency of the 1D-model, 25.15~meV, from Equation~\eqref{eq:Omega_g_1D}. 
This make sense since, by defining $E_{\rm{FC,g}}=(1/2) \sum_{\nu}(\Delta Q_{\nu})^2\omega_{\nu}^2$ , one can show that Equation~\eqref{eq:Omega_g_1D} is equivalent to Equation~\eqref{eq:Omega_multiD}. 
The computed difference between those two values gives an estimation of the error made by using phonon modes of the undoped structure. Indeed, the first value is computed with the phonon of the undoped structure while the second is computed, within the 1D-model, with the doped structure. We believe that this effect is related to the difference in bond lengths (Sr-O/N/Sr versus Eu-O/N/Sr).
This small difference reveals that the phonon modes that contributes to the electronic transitions are already present in the undoped material without the addition of the Eu doping agent. 
This result is in contrast to the result of Alkauskas \textit{et al.}~\cite{alkauskas2014first} in which the luminescent lineshape of diamond nitrogen-vacancy center could not be correctly described with the phonon modes of the undoped material.

The total Huang-Rhys factor $S_{\rm multi-D}=\sum_{\nu}S_{\nu}$ is 3.05, smaller than the one-dimensional $S_{\rm 1D}$ = 3.82. This difference can be explained by developing Eqs. \eqref{eq:S} and \eqref{eq:S_nu}. 
One shows, with the same $E_{\rm{FC,g}}$ definition, that $S_{\rm{1D}} \propto (\sum_{\nu}p_{\nu}\omega_{\nu}^2)^{1/2}$ is always larger than $S_{\rm multi-D}=\sum_{\nu}S_{\nu} \propto \sum_{\nu}p_{\nu}\omega_{\nu}$. 
$S_{\rm 1D}$ deviates largely from $S_{\rm multi-D}$ when the spread of the spectral function $S(\hbar\omega)$ is large, which is the case here.
We will see that using the Huang-Rhys factor as computed in the more precise multi-dimensional approach leads to a luminescence intensity $L(\hbar\omega)$ that better matches the experiment.

Following the generating function approach described in Section~\ref{sec:Theory}, the emission spectral function  $A(\hbar\omega)$ was calculated based on the spectral decomposition $S(\hbar\omega)$ (Equation \eqref{eq:A(hw)_generating}). The resulting curve is then convoluted with a Gaussian of fixed FWHM $w$. The normalized luminescence intensity is finally obtained by considering the cubic frequency dependence : $L(\hbar\omega)=C\omega^3A(\hbar\omega)$.  

\begin{figure}
	\centering
	\includegraphics[width=8.5cm]{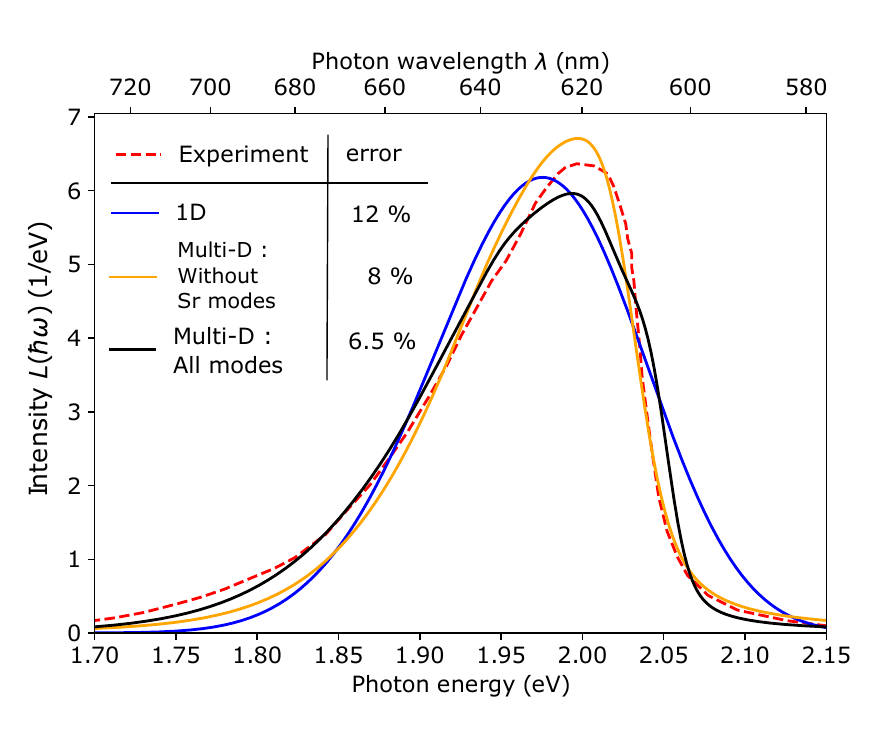}
	\caption{Comparison of normalized luminescence intensity $L(\hbar\omega)$ with experiment . Blue: 1D model, with a single effective mode. Orange:  multi-dimensional model, where the Sr-dominated modes were excluded. Black: multi-dimensional model, with a full phonon decomposition. For each curve, the broadening parameters and E$_{\rm{ZPL}}$ were optimized to minimize the error as defined in the text. The best fit is obtained within the multi-dimensional model when all phonon modes are taken into account.}
	\label{fig:Multi_D-1D}
\end{figure}

On \textbf{Figure S6}, we present the luminescence intensity as computed in this work at 0 K following 1D model, multi-D model without the Sr-dominated modes (only considering orange area of Figure \ref{fig:S_(hw)}d) and the full multi-D model with all modes. No shift of the zero-phonon line was considered and a small smearing was applied.

In order to compare these results with the experimental spectrum (15K) and to deal with the indeterminacy of the experimental $E_{\rm{ZPL}}$, we proceed in the following way.   
For each case, we adjust the shift of the curve (modification of $E_{\rm{ZPL}}$) and the gaussian and lorentzian broadening parameters ($w$ and $\gamma$) to minimize the error defined as the integral of the absolute difference between the simulated and experimental curves (in energy scale), divided by two. 
Since all spectra are normalized such that $\int L(\hbar\omega)d(\hbar\omega)=1$, this ensures an error bounded to [0 - 1]. 
The resulting curves are presented in \textbf{Figure \ref{fig:Multi_D-1D}}.
%

%
We find that, in the 1D case, using $w$ = 92~meV, $\gamma$ = 12~meV and a shift of 26~meV toward higher energies gives the best fit (error of 12 \%).
In the multi-D case without considering the Sr-dominated modes, using $w$ = 30~meV, $\gamma$ = 75~meV, and a shift of 14~meV towards lower energies gives the best fit (error of 8 \%). When using all the phonon modes in the multi-D model, using $w$ = 22~meV, $\gamma$ = 34~meV and a shift of 14 meV toward higher energies gives the best fit (error of 6.5 \%).

Indeed, the broadening parameters $\gamma$ and $w$ used refer physically to homogeneous and inhomogeneous broadening. However, in the case of SALON phosphor, no fine vibronic structure is apparent experimentally, making the exact physical determination of these broadening parameters difficult. 
Here, $w$ and $\gamma$ parameters are used as fitting parameters to match experimental emission spectrum shape. 
No physical conclusion should be drawn from the values of $\gamma$ and $w$ that gives the best fit.
We note first that the 1D model does not reproduce well the shape of the experimental spectrum.
Indeed, it has be shown that the single mode approximation is well justified for large coupling S$\gg$1 \cite{alkauskas2016}, which is not the case here. 
We have also seen that S$_{\rm 1D}$ overestimates the more precise S$_{\rm multi-D}$. 
Moreover, the spectral decomposition $S(\hbar\omega)$ is highly asymmetric around the effective frequency : an important weight on a short energy range at low energy and relatively broad distribution after 25-30 meV.
This fine feature cannot be captured well by assuming only one effective mode as done in the 1D model.  

We observe also that considering all the phonon modes, including the Sr-dominated modes, allows to obtain the best match with the experiment. In particular, the long tail of the spectrum is only well reproduced with the full computation. This underlies the importance of the long-range Sr displacements in shaping the emission spectrum of Sr[Li$_2$Al$_2$O$_2$N$_2$]:Eu$^{2+}$.
Surprisingly, the spectrum obtained without those Sr modes provides already a good estimate of the experiment. This is however fortuitous : indeed, by only considering the high-energy part of $S(\hbar\omega)$, the total Huang-Rhys factor is decreased but the effective frequency is increased. On the emission spectrum, these two effects are counter balancing each other (see Equation \eqref{eq:W_0}) which explains the small difference between orange and black curve of Figure \ref{fig:Multi_D-1D}. 

To conclude this section, we make the link between the small FWHM of the emission spectrum of SALON with its particular structure. 
A small FWHM is achieved by reducing at the same time the normal coordinate change $\Delta Q$ and the effective frequency $\hbar\Omega_g$ (see Equation \eqref{eq:W_0}). 
In the case of SALON, the small $\Delta Q$ comes from the high rigidity of the host structure. The small effective frequency, on the other hand, comes from the large coupling of the Sr channel relaxation with low-frequency phonon modes involving these channels. 
We believe that those features are not specific to SALON but are also valid for other UCr$_4$C$_4$ type phosphors that present a rigid host structure and this peculiar Ca/Sr/Ba channel. 
%

\section{Conclusions}
\label{sec:Conclusion}

In the present work, we characterize the newly discovered narrow-band emission Sr[Li$_2$Al$_2$O$_2$N$_2$]:Eu$^{2+}$ red phosphor from first principles. 
Experimental structural parameters are reproduced within 1~\% error for undoped SALON. 
We analyze the structural change of the system upon emission.
Two relaxation patterns are distinguished : a nearly isotropic expansion of the cuboid environment of Eu$^{2+}$ composed of 4 oxygen and 4 nitrogen, with small distortion and a long-range relaxation of the strontium atoms belonging to the strontium channels along c-axis. 
At the level of the electronic structure, the $\Delta$SCF method is able to correctly place the 4f and 5d states in the band gap. 
The lowest 5d state is identified as presenting a 5d$_z^{2}$ character and shows a large spatial delocalization along the Sr channel. 
Concerning the luminescent properties, we show that SALON presents a small effective displacement between excited and ground state.
By projecting the atomic displacements induced by 4f-5d transition onto the phonon eigenvectors, we identify the phonon modes that couples the most with the transition.
This spectral decomposition is dominated by a coupling to low-frequency acoustic and $B_u$ modes involving long-wavelength displacement of these Sr channels, which lowers the effective vibrational frequency. 
The high-frequency part of this decomposition originates from the coupling
to a large number of modes involving O, N and Al.
This contradicts the initial guess of Hoerder \textit{et al.}~\cite{hoerder2019sr} that argued that the isotropic structural relaxation, coming from the high symmetry environment of Eu$^{2+}$, reduces the number of different vibrational states involved in the emission process.
Overall, we computed a total Huang-Rhys parameter of $S = 3.05$. This small coupling, in combination with a moderate effective frequency, yields a small band width. 

Finally, we compute the luminescence intensity spectrum based on this phonon decomposition and compare it with the spectrum obtained with the simpler one-dimensional approach. 
With respect to experiment, we conclude that the multi-D approach is twice as accurate as the 1D approach. In particular, the tail of the spectrum is very well described within the multi-D approach.

These results are important in two aspects. First, they directly provide a theoretical understanding of the observed experimental spectrum. 
In particular, we highlighted the importance of the Sr channel oriented along c-axis. The spectrum of other UCr$_4$C$_4$-type phosphors that present similar Ca/Sr/Ba channel might be similarly affected.
Second, the success of the methodology used provides confidence in its potential applicability in the theoretical study of other important phosphors and their luminescent intensity spectra. 
It could be for instance interesting to apply this multi-dimensional approach to a low-temperature phosphor emission spectrum that still presents an apparent vibronic fine structure, allowing a further validation of the computational methodology.

\section{Computational method}
\label{sec:Comp_method}
\subsection{DFT parameters}

Calculations are performed using density-functional theory (DFT) using ABINIT~\cite{Abinit2002}. Core electrons are treated with the PAW method \cite{torrent2008implementation}.
The generalized gradient approximation (GGA-PBE) is used to treat exchange-correlation effects \cite{perdew1996generalized} and a Hubbard U term (U=7~eV) is added on $4f$ states of europium, in line with the value used in a study of fifteen Eu-doped materials based on the same PAW atomic dataset~\cite{jia2017first}, thus allowing the consistent comparison with these results. 

Calculations on europium-doped SALON are conducted with a supercell technique. 
Most results shown in this manuscript are computed with a 1$\times$1$\times$12 supercell composed of 216 atoms. Since there are 24 Sr atoms in this supercell, this corresponds to a theoretical doping concentration of 4.16\% while the experimental one is 0.7\%. Still, since our methodology does not account for concentration-dependent effects such as self-absorption or energy transfer, our only concern is the spurious interaction between Europium periodic replica, that should be numerically controlled. 
To this aim, we performed a careful convergence study of our main results with respect to the supercell size, that can be found in the supplementary material.
Because of the anisotropy of the material (tetragonal structure), we checked both an increase of the size in the x-y plane and an increase along z axis. %
It appears that the critical parameter to converge is the long range structural relaxation of the strontium
atoms along z-axis, explaining the particular elongated shape of our supercell. %

For all supercell calculations, the structures are relaxed below a maximal residual force of $10^{-5}$~Hartree/Bohr. The cut-off kinetic energy is 30~Ha and a 2$\times$2$\times$2 wavevector grid is used. 
Phonons are calculated for the undoped structures at the $\Gamma$ point. 

Although spin-orbit coupling can present a noticeable effect on the excited state electronic configuration, it is not considered here due to its small effect on the emission spectra.
This work focuses on the lowest excited state of the $4f^65d^1$ multiplet ($^7F_J$, J = 0-6) splitting and does not try to account for the other states.

\subsection{Simulating the Eu 5d excited state}

The treatment of the Eu $4f^65d^1$ excited state and the resulting Eu $5d$ electron - $4f$ hole interaction might rely on the Bethe-Salpeter equation~\cite{onida2002electronic} but this would be too computationally demanding for such supercell system. 
For this reason, we rely on the $\Delta$SCF method whereby the eigenfunction associated to the highest Eu $4f$ state is forced to be unoccupied while the next energy state (identified as a $5d_z^2$ state) is constrained to be occupied. Transition energies are then computed as difference of two total energies. 
The electron-hole interaction is here mimicked through the promotion of the Eu 4$f$ electron to the Eu 5$d$ state. By doing that, the seven $4f$ levels split into one unoccupied band that stays within the gap, and six occupied bands that shift downwards, hybridizing inside the valence band~\cite{jia2017first}. 
We suppose that the 4f-5d transition  only affects the local geometry of the crystal and not the macroscopic state of the crystal, meaning that the lattice parameters of the crystal will not change. 
Thus, atomic relaxation in the excited state is done at fixed lattice parameters obtained in the ground state.

Additional details on the $\Delta$SCF method are given in supplementary informations.  
%

\medskip
\textbf{Acknowledgements} \par 
We thank a referee for a remark about the convergence study, that lead us to reconsider the domination of the cuboid distortion in the spectral function. Computational resources have been provided by the CISM/UCLouvain and the CECI funded by the FRS-FNRS Belgium under Grant No. 2.5020.11, as well as the Tier-1 supercomputer of the 
F\'ed\'eration Wallonie-Bruxelles, infrastructure funded by the Walloon Region under the grant agreement No. 1117545. J.B. acknowledges support from the Communauté française de Belgique through the SURFASCOPE project (ARC 19/24-057). Y.J. has been supported by the FRS-FNRS Belgium through a Chargé de
recherches fellowship.
S.P. acknowledge support from the European Unions Horizon 2020 Research and Innovation Programme, under the Marie Sk\l{}odowska-Curie Grant Agreement SELPH2D No.~839217.
\medskip


\end{document}


\beginsupplement
\noindent
{\Large Supporting informations}
\vspace{1cm}

\noindent
\textbf{\Large Importance of long-range channel Sr displacements for the narrow emission in Sr[Li$_2$Al$_2$O$_2$N$_2$]:Eu$^{2+}$phosphor}
\vspace{0.5cm}

\noindent
\textit{Julien Bouquiaux *, Samuel Ponc\'{e}, Yongchao Jia, Anna Miglio, Masayoshi Mikami, Xavier Gonze}


\section{Details on the $\Delta$SCF method}

The constrained-occupation $\Delta$SCF method, by which an excitation energy is computed as difference of two total energies, has been used in different contexts. 
It started in atomic physics or core-level spectroscopy, see~\cite{Hedin1969b} and references therein, and was based on the Hartree-Fock total energies at that time.
Concerning density-functional theory, in 1976, Gunnarson and Lundqvist~\cite{Gunnarsson1976,Jones1989} proved a theorem stating that the DFT total energy of the ground-state of a particular (constrained) symmetry (or irreducible representation) can be exactly obtained.
However, the DFT exchange-correlation functional
to be used is not the one of the unconstrained ground-state: a specifically taylored exchange-correlation functional should be used in the context of such exact theorem.

However, to our knowledge no study of realistic materials has relied on such a specialized exchange-correlation functional, but all used the (unconstrained) ground-state one, as we are doing. 
Indeed the usage of the technique is pragmatic, and without formal justification.
The results are nevertheless qualitatively and quantitatively helpful~\cite{Herbst1978, Tozer2000, Hellman2004, Gavnholt2008}.
In the context of the luminescence 
of phosphors, this methodology has been successfully applied in references~\cite{Marsman2000,chaudhry2011first,chaudhry2014first,jia2016first,jia2017first,jia2017assessment,Jia-2018,Jia-2019-1,Jia-2019-2} to Ce$^{3+}$ and Eu$^{2+}$ inorganic oscillators compounds.
In particular, Jia \textit{et al.}~\cite{jia2017assessment} estimated the predictive power of this method for fifteen representatives Eu$^{2+}$ doped phosphors
using settings similar to ours.
They obtained an absolute error below 0.3~eV~\cite{jia2017first} with respect to experiment on absorption and emission energies.

In order to avoid confusion, let us specify that none of the eigenfunctions within the PAW method are pure states of a given atomic orbital symmetry. 
They are obtained from the PAW hamiltonian, in which linear combinations of projector-augmented planewaves describe states originating from different atomic orbitals on different atomic sites or from the space between atomic sites. 
So, when we refer to ``Eu 4$f$" or ``Eu 5$d$" states, or the likes, we actually intent to designate the eigenfunctions that
are predominantly of ``Eu 4$f$" or ``Eu 5$d$" characteristics, or the likes. 
They are never pure ``Eu 4$f$" or ``Eu 5$d$" states.

\section{Supercell convergence study}

\begin{figure}[h!]
    \centering
    \includegraphics[width=0.6\linewidth]{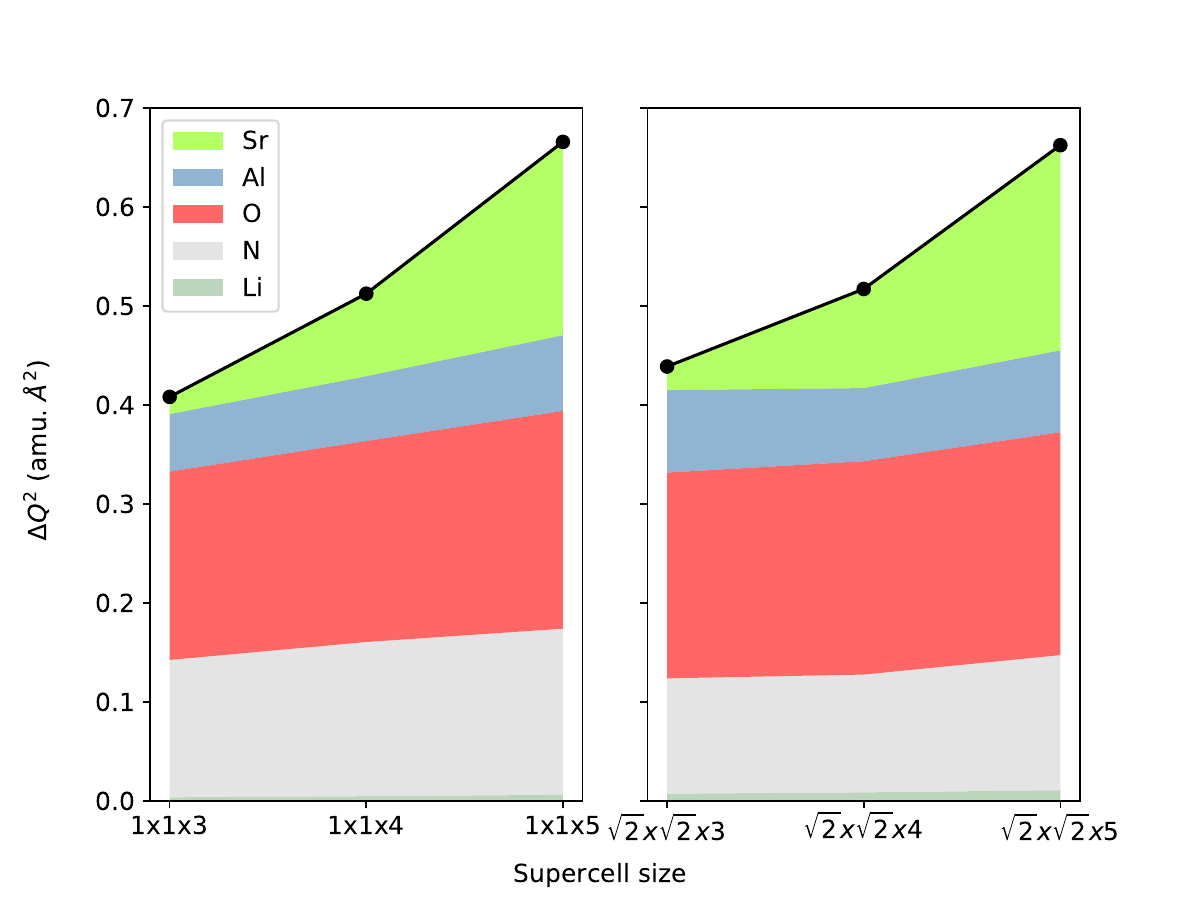}
    \caption{Convergence of the normal coordinate change $(\Delta Q)^2$ decomposed by atom's type as a function of the supercell size $\sqrt{2}\times\sqrt{2}\times$ N compared to the $1\times1\times$ N case. It appears that doubling the supercell in the x-y plane does not affect significantly the result.}
    \label{fig:conv_Q_sq2}
\end{figure}

\begin{figure}[h!]
    \centering
    \includegraphics[width=\textwidth]{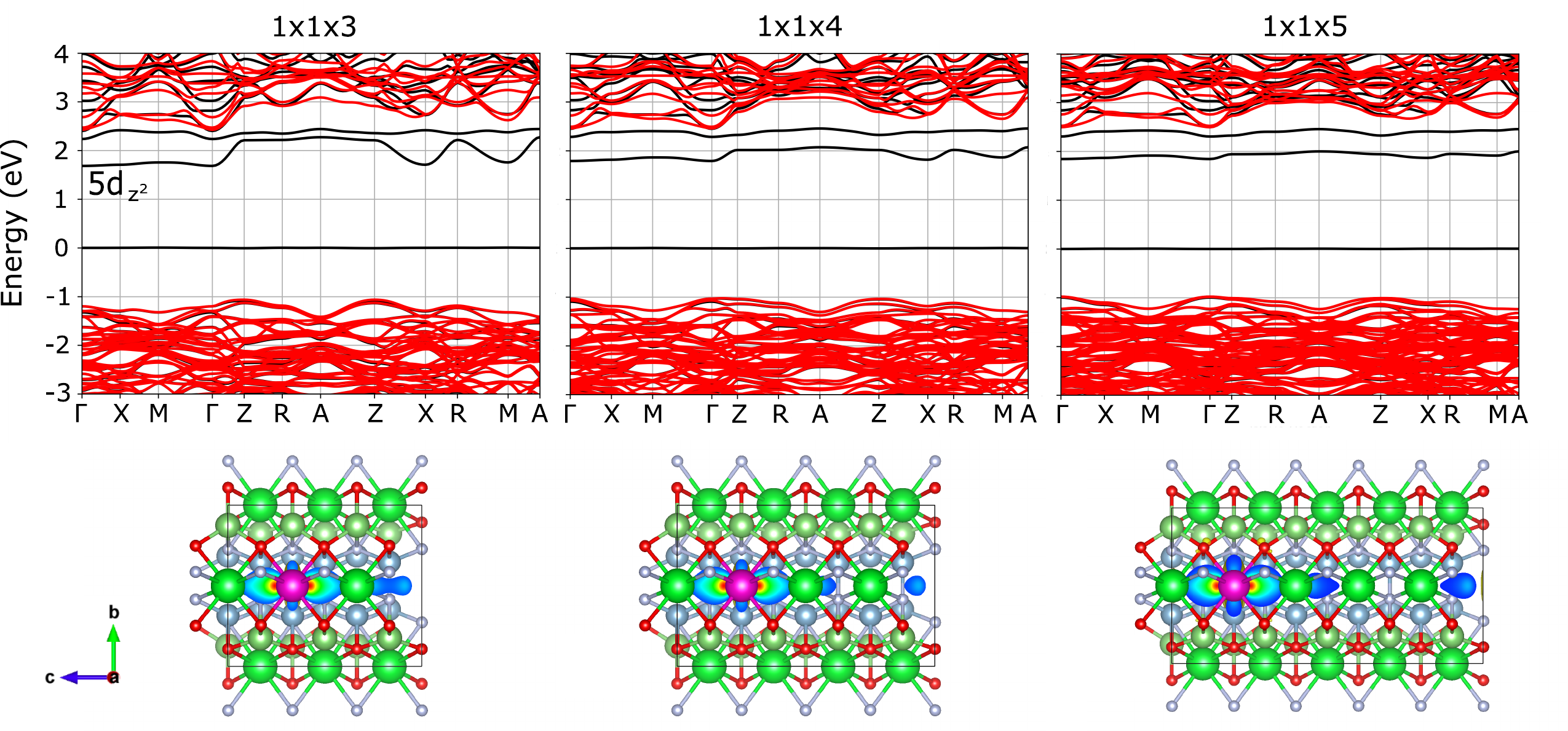}
    \caption{Up : Electronic band structure of doped SALON in its relaxed excited state $\Lambda_e^*$ for 1$\times$1$\times$3, 1$\times$1$\times$4 and 1$\times$1$\times$5 supercell composed of respectively 54, 72 and 90 atoms. The 5d$_{z^2}$ band presents a decreasing dispersive behaviour along $\Gamma$-Z with an increase of the supercell size along c-axis. Bottom : section representation of the Kohn-Sham orbitals with 5d$_{z^2}$ character.}
    \label{fig:BS_conv_supercell}
\end{figure}

\begin{figure}[h!]
	\centering
	\includegraphics[width=0.75\columnwidth]{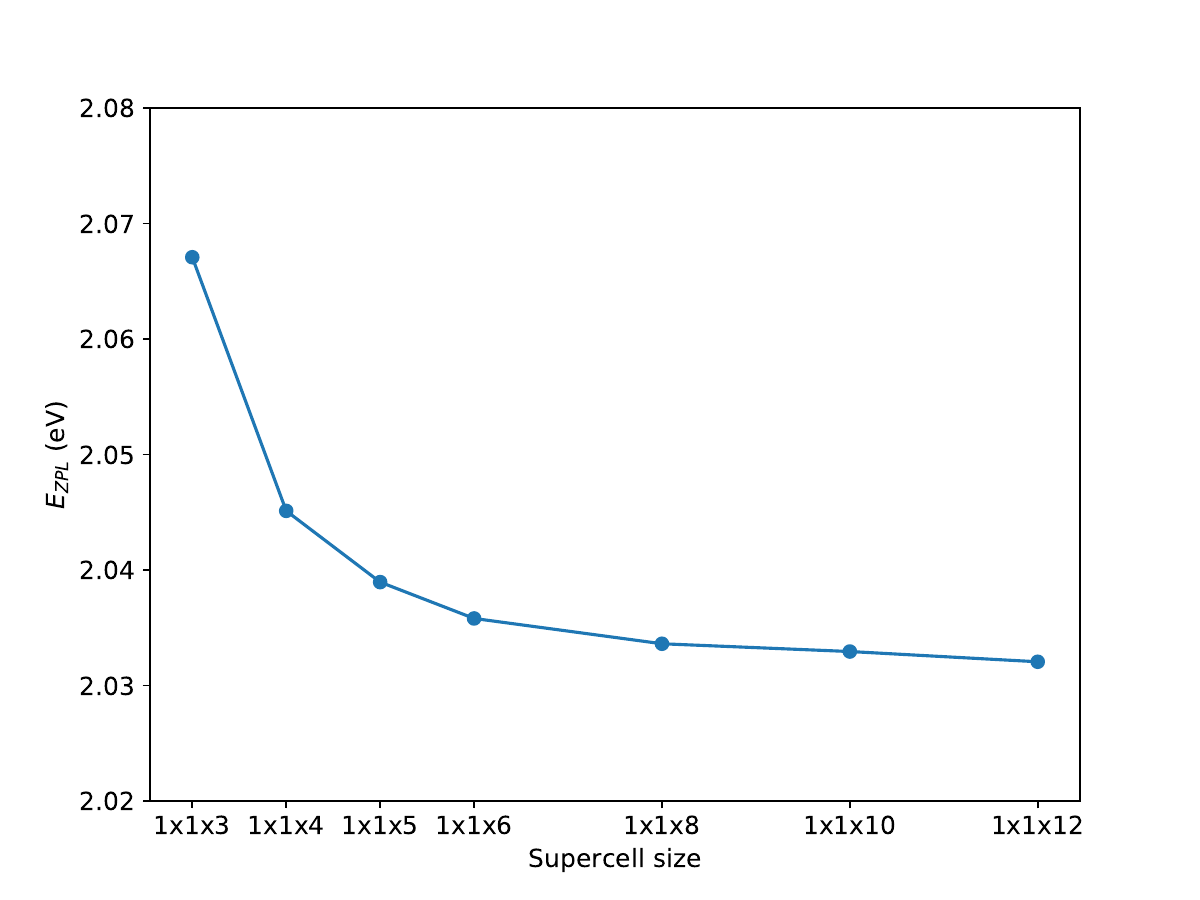}
	\caption{Convergence of the zero-phonon line energy E$_{ZPL}$ as a function of the supercell size. As opposed to the structural relaxation parameters, the values associated to small sizes are already reasonably good. }
	\label{fig:E_zpl_conv}
\end{figure}

\begin{figure}[h!]
	\centering
	\includegraphics[width=\columnwidth]{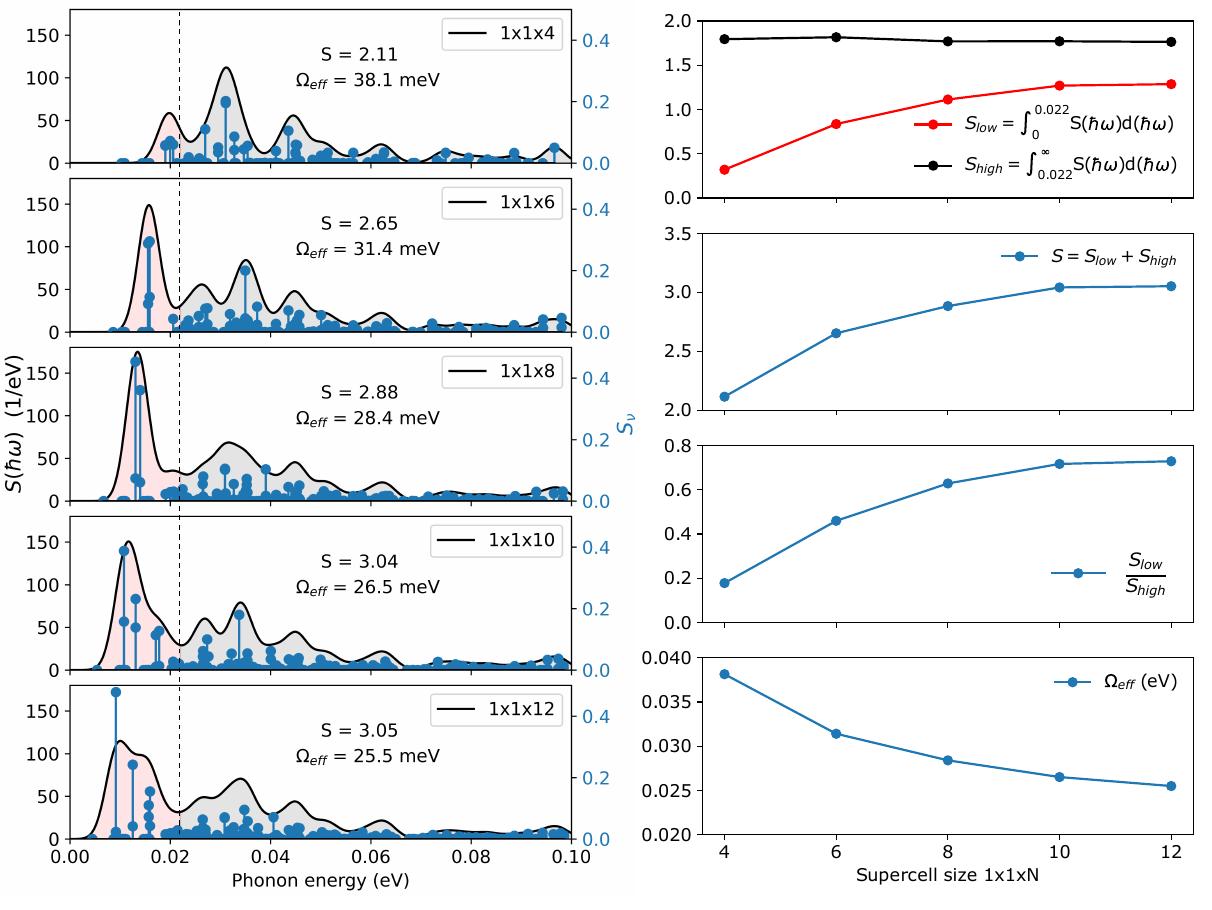}
	\caption{Convergence of the spectral function $S(\hbar\omega)$ as a function of the supercell size.}
	\label{fig:S(hw)_conv}
\end{figure}

We performed a carefull convergence study of our main results with respect to the supercell size. 
\newline
\newline
Concerning the critical normal coordinate change $(\Delta Q)^2$, \textbf{Figure \ref{fig:conv_Q_sq2}} shows that an increase of the size along the x-y plane does not affect the result. We can thus safely focus on an increase of the size along the tetragonal c-axis. The convergence of $(\Delta Q)^2$ with 1$\times$1$\times$N supercells is described in the main manuscript. 
\newline
\newline
We checked also how the Kohn-Sham band structure changes as we increase the supercell size 1$\times$1$\times$N. \textbf{Figure \ref{fig:BS_conv_supercell}} shows that for too small size, there exists a large dispersive behavior of the 5d$_{z^2}$ that comes from the interaction between periodically arranged europium.  
\newline
\newline
\textbf{Figure \ref{fig:E_zpl_conv}} shows the convergence of the zero-phonon line energy. It can be seen that the result is already well converged for small sizes, as opposed to $\Delta Q$. 
\newline
\newline
\textbf{Figure \ref{fig:S(hw)_conv}} provides the convergence of the spectral function of electron-phonon coupling. Integrated contribution from high-energy phonons (threshold fixed at 22 meV) is nearly constant with the supercell size while integrated contribution from low-energy phonons slowly converges. Their sum and their ratio is nearly converged for a 1x1x12 supercell. On the other hand, the exact shape of the low-energy part is still not obtained. This issue, due to the coupling to long-wavelength acoustic modes, was already observed by Alkauskas \textit{et. al} for NV center in diamond in \cite{alkauskas2014first} and recently further analysed in \cite{razinkovas2020vibrational}.

\begin{figure}[h!]
	\centering
	\includegraphics[width=\columnwidth]{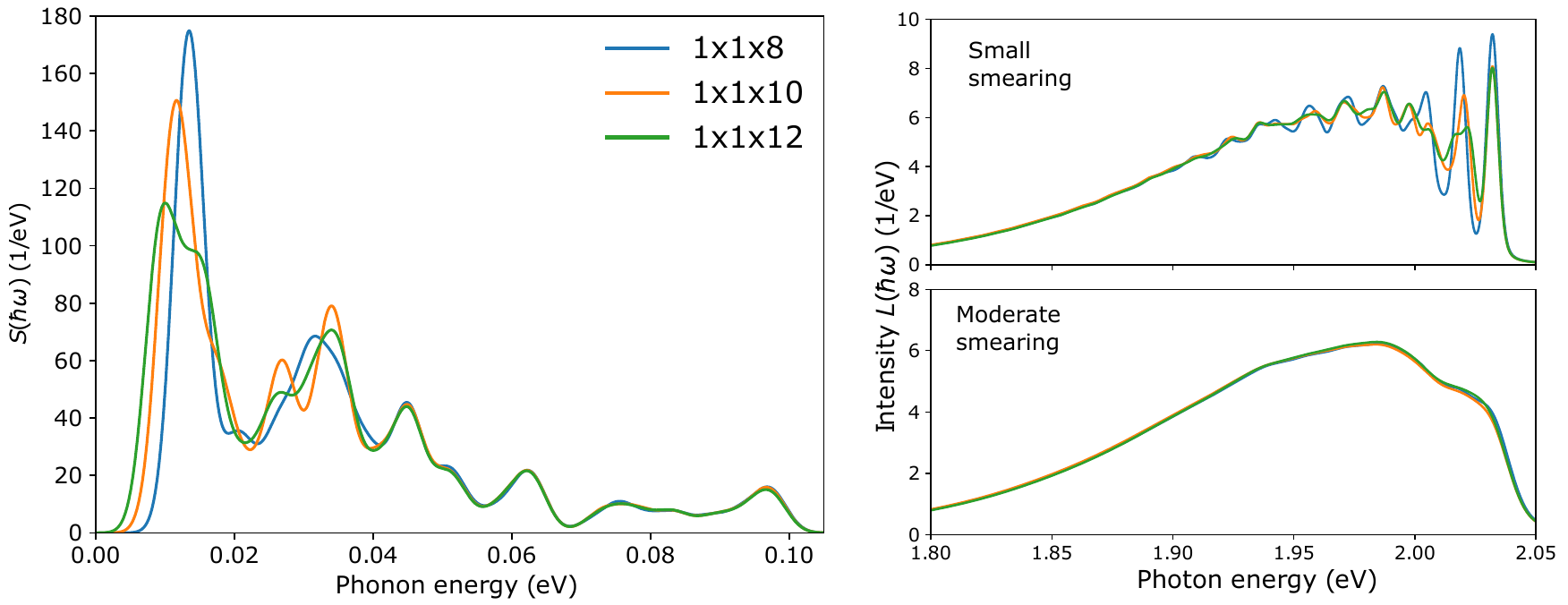}
	\caption{Spectral function $S(\hbar\omega)$ and luminescence intensity $L(\hbar\omega)$ for the three largest simulated supercells.}
	\label{fig:conv_S(hw)_L(hw))}
\end{figure}

As it can be seen on \textbf{Figure \ref{fig:conv_S(hw)_L(hw))}}, this slow convergence of the low-energy shape of $S(\hbar\omega)$ leads to small deviations in the non-smeared emission spectrum near the zero-phonon line. However, already with moderate smearing, the emission spectrum is converged for the 1x1x8 supercell. Since the experimental spectrum does not show defined vibronic structure that we could  compare with our theoretical calculation, we stick to our 1x1x12 supercell calculations and do not attempt to increase further the supercell size. 

\begin{figure}[h!]
	\centering
	\includegraphics[width=0.7\columnwidth]{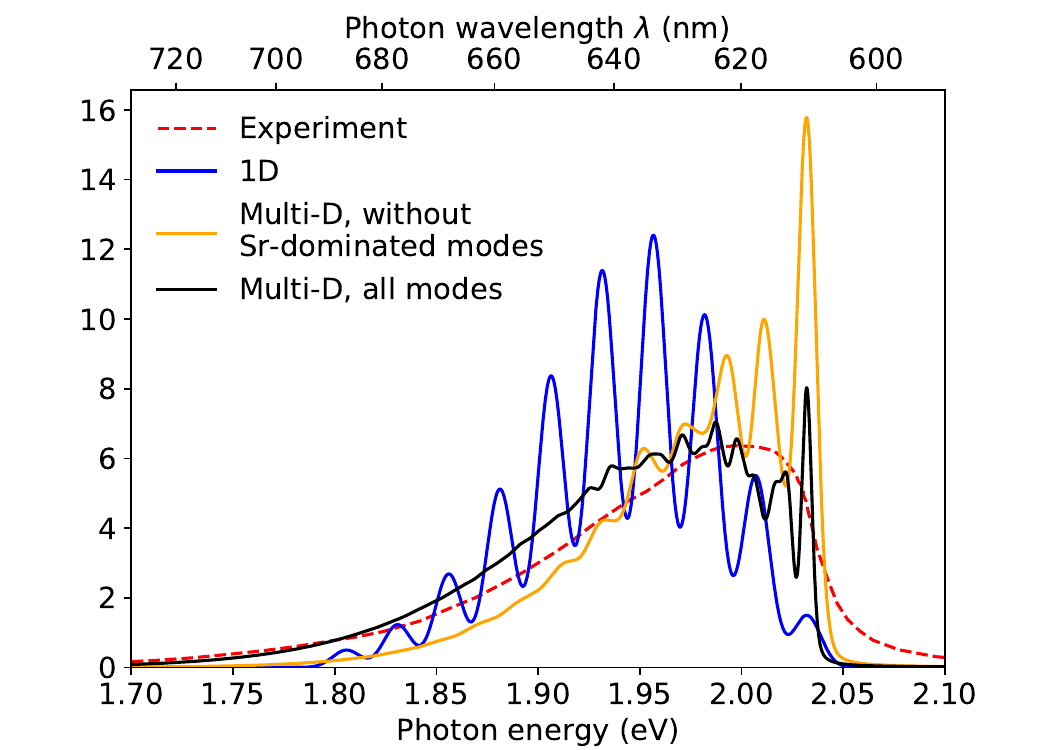}
	\caption{Comparison of normalized luminescence intensity $L(\hbar\omega)$ with a small smearing and without shift of E$_{\rm{ZPL}}$ . Blue: 1D model, with a single effective mode. Orange:  multi-dimensional model, where the Sr-dominated modes were excluded. Black: multi-dimensional model, with a full phonon decomposition. We notice how the multi-D model reproduce well the long tail of the experimental spectrum. }
	\label{fig:Emission_spectra_no_smear}
\end{figure}

\clearpage

\section{Vibrational properties}

\begin{figure}[h!]
	\centering
	\includegraphics[width=0.7\textwidth]{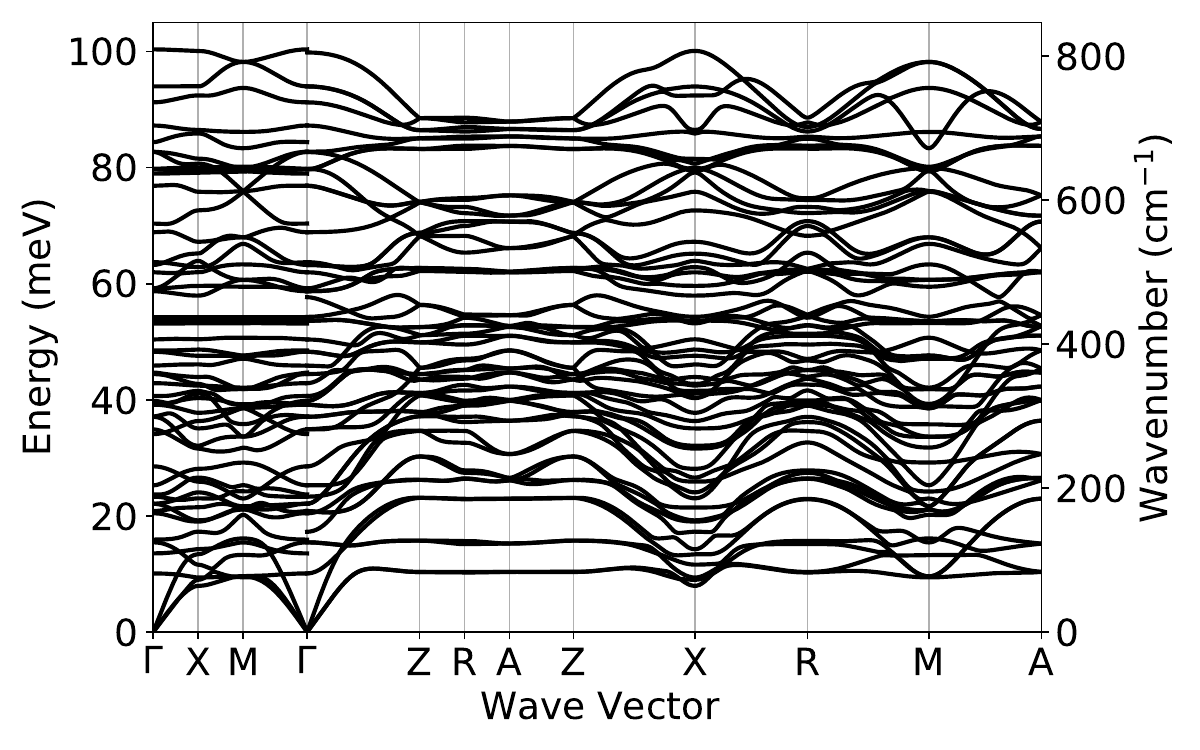}
	\caption{Phonon band structure of undoped SALON computed using DFPT for a primitive cell composed of 18 atoms and a 2$\times$2$\times$6 Monkhorst-Pack wavevector grid.}
	\label{fig:Phonon_BS}
\end{figure}

\begin{figure}[h!]
    \centering
    \includegraphics[width=0.6\columnwidth]{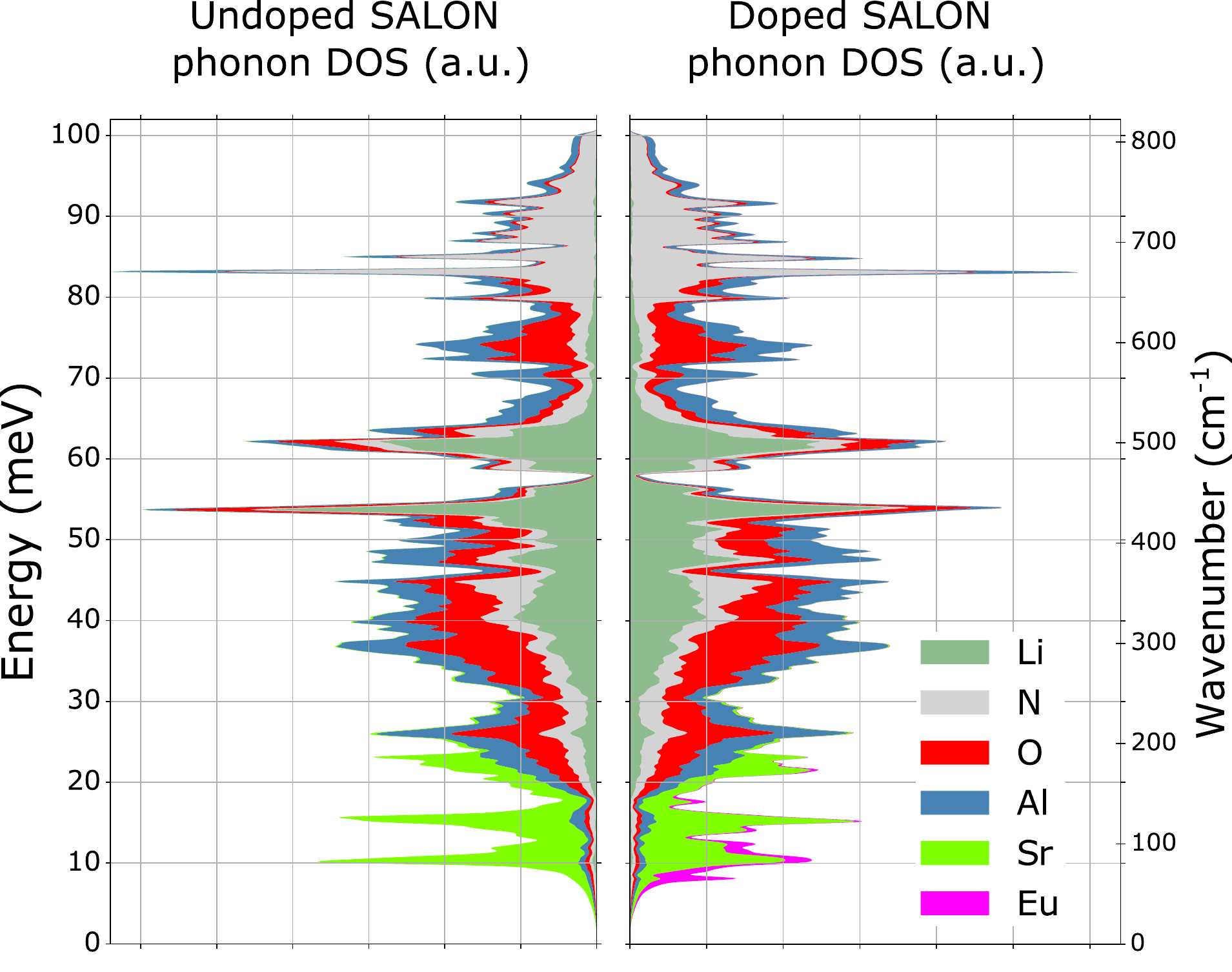}
	\caption{\label{fig:Phonon_DOS_comp}Comparison between the phonon DOS of undoped SALON and europium-doped SALON.}
\end{figure}

We provide in this section the vibrationnal properties of SALON as computed in this work.
\textbf{Figure \ref{fig:Phonon_BS}} shows the complete phonon band structure of undoped SALON. 
We provide on \textbf{Figure \ref{fig:Phonon_DOS_comp}} a comparison between the phonon DOS of the undoped/Eu-doped structure for a  54-atoms supercell. Small deviations are observed in the low-energy region dominated by Sr-based modes. It is expected that those differences are further reduce with larger supercells.  

The phonon frequencies at the zone center and the irreducible representation of each mode are given in \textbf{Table \ref{tab:PhononFreq}}.  For each mode, the associated irreducible representation and frequency are given. Letters $A$ and $B$ are for non-degenerate (single) modes where we have only one set of atom vector displacement for a given wavenumber $\omega$. $A$ indicates a symmetry with respect to the principal rotation axis (z here). $B$ indicates an anti-symmetry with respect to the principal rotation axis. Letter $E$ refers to doubly degenerated modes (two sets of atom vector displacements for a given wavenumber $\omega$). The subscripts $g$ and $u$ refer to a mode that is respectively symmetric or anti-symmetric to inversion. Among the 54 modes (18 atoms in 3 dimensions), we observe 8 $A_g$ modes, 5 $A_u$ modes, 8 $B_g$ modes, 5 $B_u$ modes, 4 doubly degenerated $E_g$ modes and 10 doubly degenerated $E_u$ modes.

Born effective charges, low and high-frequency dielectric tensor were computed with a dense 4$\times$4$\times$12 k-point grid to ensure converged values. The Born effective charges can be found in \textbf{table \ref{tab:Born}}.
The electronic contribution to the macroscopic dielectric tensor is found to be ($\epsilon_{xx}^\infty, \epsilon_{yy}^\infty$, $\epsilon_{zz}^\infty$) = (3.966, 3.966, 4.041). For comparison, the dielectric constants of the oxide red phosphor SrO:Eu and the nitride red phosphor Sr$_2$SiN$_8$ are respectively (3.76, 3.76, 3.76) and (4.21, 4.60, 4,50)~\cite{mikami20105d}.
The low-frequency dielectric tensor includes the response of the ions, whose motion will be triggered by the electric field, with values $\epsilon_{xx}^0=\epsilon_{yy}^0$ = 11.974, $\epsilon_{zz}^0$ = 13.048.

\begin{table}[h!]
	\begin{tabular}{|p{2cm}|p{2cm}|p{2cm}|p{2cm}|p{2cm}|p{2cm}|}
		\hline
		\multicolumn{2}{|p{4cm}|}{\textbf{Phonons frequencies at $\Gamma$ ($cm^{-1}$) - Irreducible representation}} & \multicolumn{2}{p{4cm}|}{\textbf{Phonons frequencies at $\Gamma$ ($cm^{-1}$), direction $\Gamma - X$ (0 1 0)}} & \multicolumn{2}{p{4cm}|}{\textbf{Phonons frequencies at $\Gamma$ ($cm^{-1}$), direction $\Gamma - Z$ (0 0 1)}} \\ \hline
		-0.01  $E_u$   & 391.51 $E_u$   & -0.01 $E_u$     & \textbf{391.51} $E_u$    & -0.01 $E_u$   & 391.51  $E_u$  \\ \hline
		-0.01  $E_u$   & 391.51 $E_u$   & -0.01 $E_u$    & \textbf{393.36}  $E_u$    & -0.01  $E_u$   & 391.51  $E_u$  \\ \hline
		0.00   $A_u$   & 407.33 $A_g$   & 0.00 $A_u$    & 407.33 $A_g$    & 0.00 $A_u$      & 407.33 $A_g$    \\ \hline
		82.00 $B_u$   & 424.91 $A_u$   & 82.00 $B_u$     & 424.91 $A_u$    & 82.00 $B_u$    & 428.80 $E_g$    \\ \hline
		110.02 $A_u$  & 428.80 $E_g$   & 110.02 $A_u$   & 428.80 $E_g$   & 124.72  $E_u$   & 428.80 $E_g$    \\ \hline
		124.72 $E_u$   & 428.80 $E_g$   & \textbf{124.72} $E_u$   & 428.80 $E_g$   & 124.72  $E_u$   & 434.58 $B_u$   \\ \hline
		124.72 $E_u$    & 434.58 $B_u$   & \textbf{128.74} $E_u$   & 434.58 $B_u$   & \textbf{140.29} $A_u$    & \textbf{462.48} $A_u$    \\ \hline
		166.15 $B_g$    & 472.14 $E_u$   & 166.15 $B_g$  & \textbf{472.14} $E_u$      & 166.15 $B_g$    & 472.14 $E_u$   \\ \hline
		169.03 $E_u$    & 472.14 $E_u$   & \textbf{169.03} $E_u$  & \textbf{476.57} $E_u$    & 169.03 $E_u$   & 472.14 $E_u$   \\ \hline
		169.03 $E_u$    & 477.39 $A_g$   & 179.25 $E_g$   & 477.39 $A_g$    & 169.03 $E_u$   & 477.39 $A_g$     \\ \hline
		179.25 $E_g$    & 497.27 $B_g$   & 179.25 $E_g$   & 497.27 $B_g$    & 179.25 $E_g$    & 497.27 $B_g$    \\ \hline
		179.25 $E_g$    & 510.37 $E_u$   & 188.13 $A_g$    & \textbf{510.37} $E_u$   & 179.25 $E_g$    & 510.37 $E_u$   \\ \hline
		188.13 $A_g$    & 510.37 $E_u$   & \textbf{192.00} $E_u$    & 517.97 $A_g$      & 188.13 $A_g$     & 510.37 $E_u$   \\ \hline
		205.37 $B_g$    & 517.97 $A_g$   & 205.37 $B_g$    & 555.00 $B_g$     & 205.37 $B_g$    & 517.97 $A_g$    \\ \hline
		231.08 $B_u$    & 555.00 $B_g$   & 231.08 $B_u$    & \textbf{569.16} $E_u$    & 231.08 $B_u$    & 555.00 $B_g$    \\ \hline
		276.46 $A_u$    & 620.09 $E_u$   & 276.46 $A_u$    & \textbf{620.09} $E_u$   & \textbf{283.30} $A_u$     & 620.09 $E_u$   \\ \hline
		284.19 $B_g$    & 620.09 $E_u$   & 284.19 $B_g$    & 637.97 $A_u$    & 284.19 $B_g$    & 620.09 $E_u$   \\ \hline
		300.53 $E_u$    & 637.97 $A_u$   & \textbf{300.53} $E_u$    & 644.66 $E_g$    & 300.53 $E_u$   & 644.66 $E_g$   \\ \hline
		300.53 $E_u$    & 644.66 $E_g$   & \textbf{301.06} $E_u$    & 644.66 $E_g$     & 300.53  $E_u$    & 644.66 $E_g$   \\ \hline
		317.55 $E_g$    & 644.66 $E_g$   & 317.55 $E_g$    & 645.55 $B_u$    & 317.55 $E_g$    & 645.55 $B_u$  \\ \hline
		317.55 $E_g$    & 645.55 $B_u$   & 317.55  $E_g$   & 667.69 $A_g$   & 317.55  $E_g$   & 667.69 $A_g$    \\ \hline
		323.22 $E_u$    & 667.69 $A_g$  & \textbf{323.22} $E_u$    & 669.15 $B_g$   & 323.22 $E_u$   & 669.15 $B_g$    \\ \hline
		323.22 $E_u$    & 669.15 $B_g$   & 327.75 $A_g$    & \textbf{680.00} $E_u$   & 323.22  $E_u$   & 705.31 $A_g$   \\ \hline
		327.75 $A_g$    & 705.31 - $A_g$   & 347.13 $B_u$   & 705.31 $A_g$    & 327.75 $A_g$    & 738.11 $B_g$    \\ \hline
		347.13 $B_u$    & 738.11 $B_g$  & 358.13 $A_g$   & 738.11 $B_g$    & 347.13 $B_u$   & 759.55 $E_u$   \\ \hline
		358.13 $A_g$    & 759.55 $E_u$   & \textbf{361.79} $E_u$    & \textbf{759.55 } $E_u$    & 358.13 $A_g$    & 759.55  $E_u$   \\ \hline
		371.75 $B_g$    & 759.55 $E_u$   & 371.75 $B_g$    & \textbf{811.27} $E_u$   & 371.75 $B_g$ &\textbf{ 807.21} $A_u$ \\ \hline
	\end{tabular}
	\caption{Phonon frequencies at gamma in $cm^{-1}$ and irreducible representation. The first column gives the frequencies when non-analytical contributions are ignored (TO modes only). The second column gives the frequencies at $\Gamma$ along (0 1 0) direction where we note the splitting of the doubly degenerate $E_u$ modes (in bold). The third column gives the frequencies at $\Gamma$ along (0 0 1) direction where we note the shifts of $A_u$ modes (in bold). }
	\label{tab:PhononFreq}
\end{table}

\begin{table}[!h]
	\centering
	\begin{tabular}{l|l|l}
		\textbf{Li1 and Li2} & \textbf{Li3 and Li4} & \textbf{N1 and N2} \\ \hline
		$\begin{pmatrix}
			1.20 & 0.02 & 0 \\
			0.04 & 0.99 & 0 \\
			0   & 0   & 0.89
		\end{pmatrix}$
		&  
		$\begin{pmatrix}
			0.99 & -0.04 & 0 \\
			-0.02 & 1.20 & 0 \\
			0   & 0   & 0.89
		\end{pmatrix}$
		&
		$\begin{pmatrix}
			-2.35 & 0.11 & 0 \\
			0.16 & -2.27 & 0 \\
			0   & 0   & -2.88
		\end{pmatrix}$ \\
		\hline
		\textbf{N3 and N4} & \textbf{O1 and O2} & \textbf{O3 and O4} \\
		\hline
		$\begin{pmatrix}
			-2.27 & -0.16 & 0 \\
			-0.11 & -2.35 & 0 \\
			0   & 0   & -2.88
		\end{pmatrix}$
		&  
		$\begin{pmatrix}
			-1.89 & 0.06 & 0 \\
			0.04 & -2.57 & 0 \\
			0   & 0   & -1.88
		\end{pmatrix}$
		&
		$\begin{pmatrix}
			-2.57 & -0.04 & 0 \\
			-0.06 & -1.89 & 0 \\
			0   & 0   & -1.88
		\end{pmatrix}$ \\
		\hline
		\textbf{Al1 and Al2} & \textbf{Al3 and Al4} & \textbf{Sr1} \\
		\hline
		$\begin{pmatrix}
			1.99 & 0.11 & 0 \\
			0.04 & 2.33 & 0 \\
			0   & 0   & 2.58
		\end{pmatrix}$
		&  
		$\begin{pmatrix}
			2.33 & -0.04 & 0 \\
			-0.11& 1.99 & 0 \\
			0   & 0   & 2.58
		\end{pmatrix}$
		&
		$\begin{pmatrix}
			2.04 & 0.11 & 0 \\
			0.16 & 3.11 & 0 \\
			0   & 0   & 2.58
		\end{pmatrix}$ \\
		\hline
		\textbf{Sr2} &  &  \\
		\hline
		$\begin{pmatrix}
			3.11 & -0.16 & 0 \\
			-0.11 & 2.04 & 0 \\
			0   & 0   & 2.58
		\end{pmatrix}$ &&
		
	\end{tabular}
	\caption{Born effective charge tensor of each atom}
	\label{tab:Born}
\end{table}

\clearpage
\section{Additional figures}

\begin{figure}[h!]
	\centering
	\includegraphics[width=0.5\columnwidth]{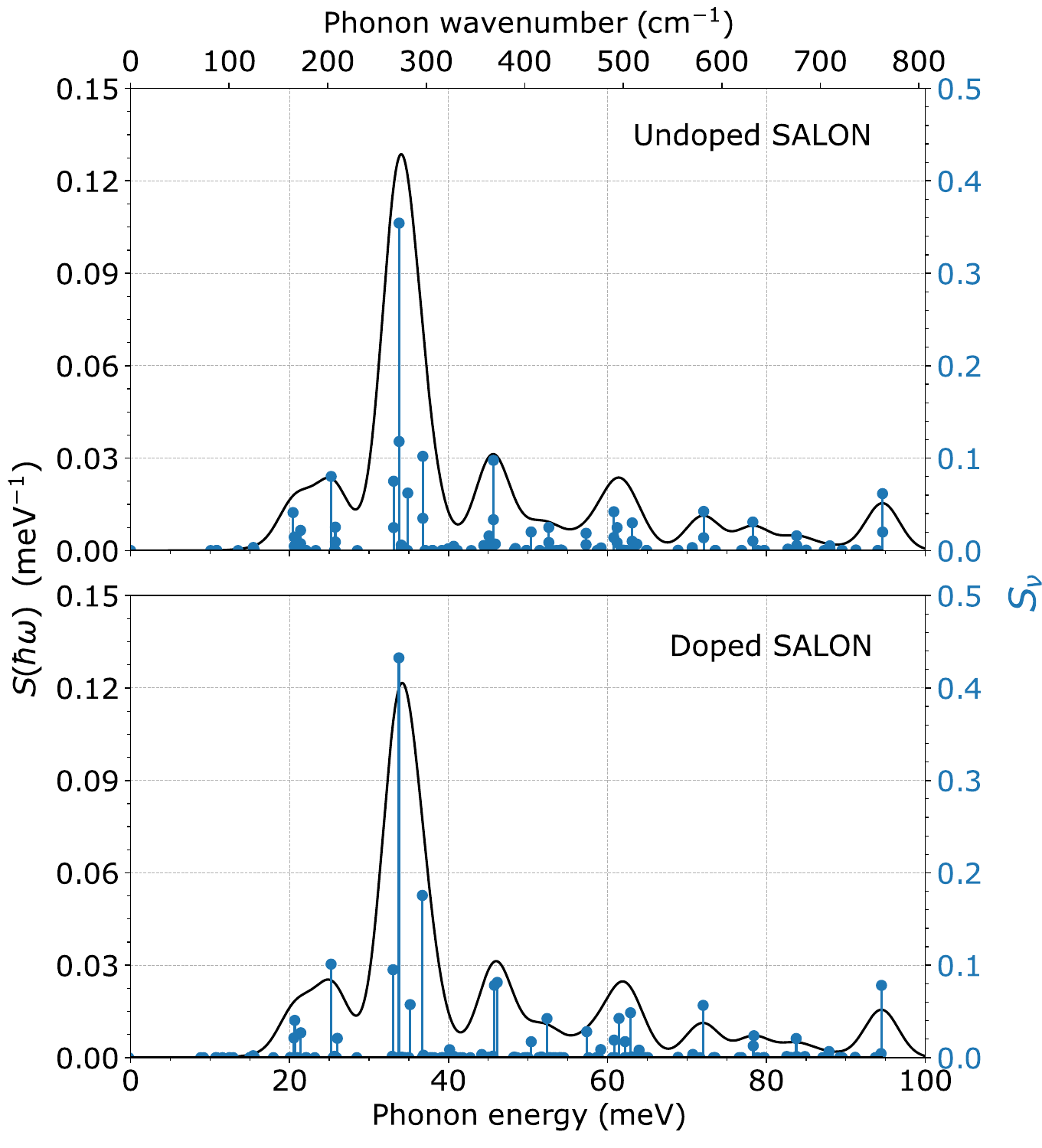}
	\caption{Partial Huang-Rhys factors $S_{\nu}$ and spectral decomposition $S(\hbar\omega)$  using a 5~meV Lorentzian broadening with the phonon modes of undoped SALON (up) and Eu-doped SALON (bottom). Computed with a 1x1x3 supercell. The difference being negligible, we infer that using the phonons modes of the undoped structure for larger supercell calculations is justified.}
	\label{fig:S_(hw)}
\end{figure}

\begin{figure}[h!]
    \centering
    \includegraphics[width=0.6\linewidth]{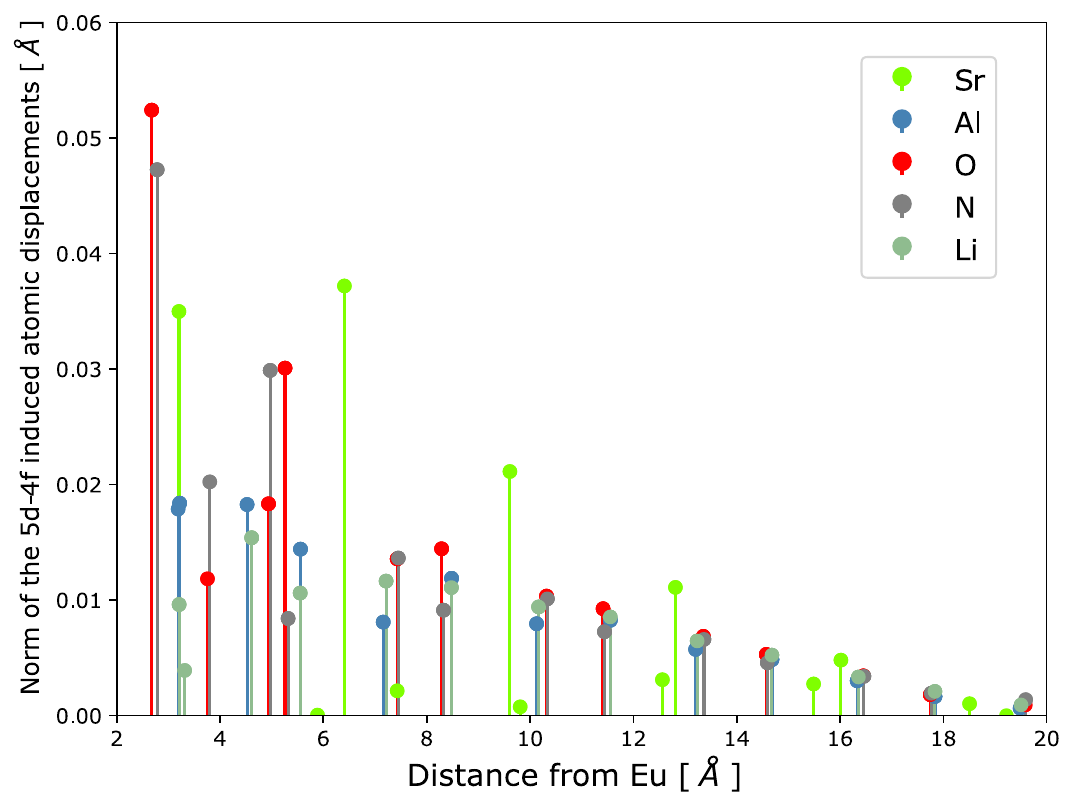}
    \caption{Norm of the 5d-4f atomic displacements as a function of the distance from europium. The first oxygen and nitrogen peaks comes from the expansion of the nearly cubic polyhedron surrounding the Eu. The Sr channel relaxation is also apparent. We recall that those values are multiplied by the respective atomic masses before used in the computation of the luminescent properties, explaining the importance of Sr atoms. }
    \label{fig:}
\end{figure}

\begin{figure}[h!]
    \centering
    \includegraphics[width=\linewidth]{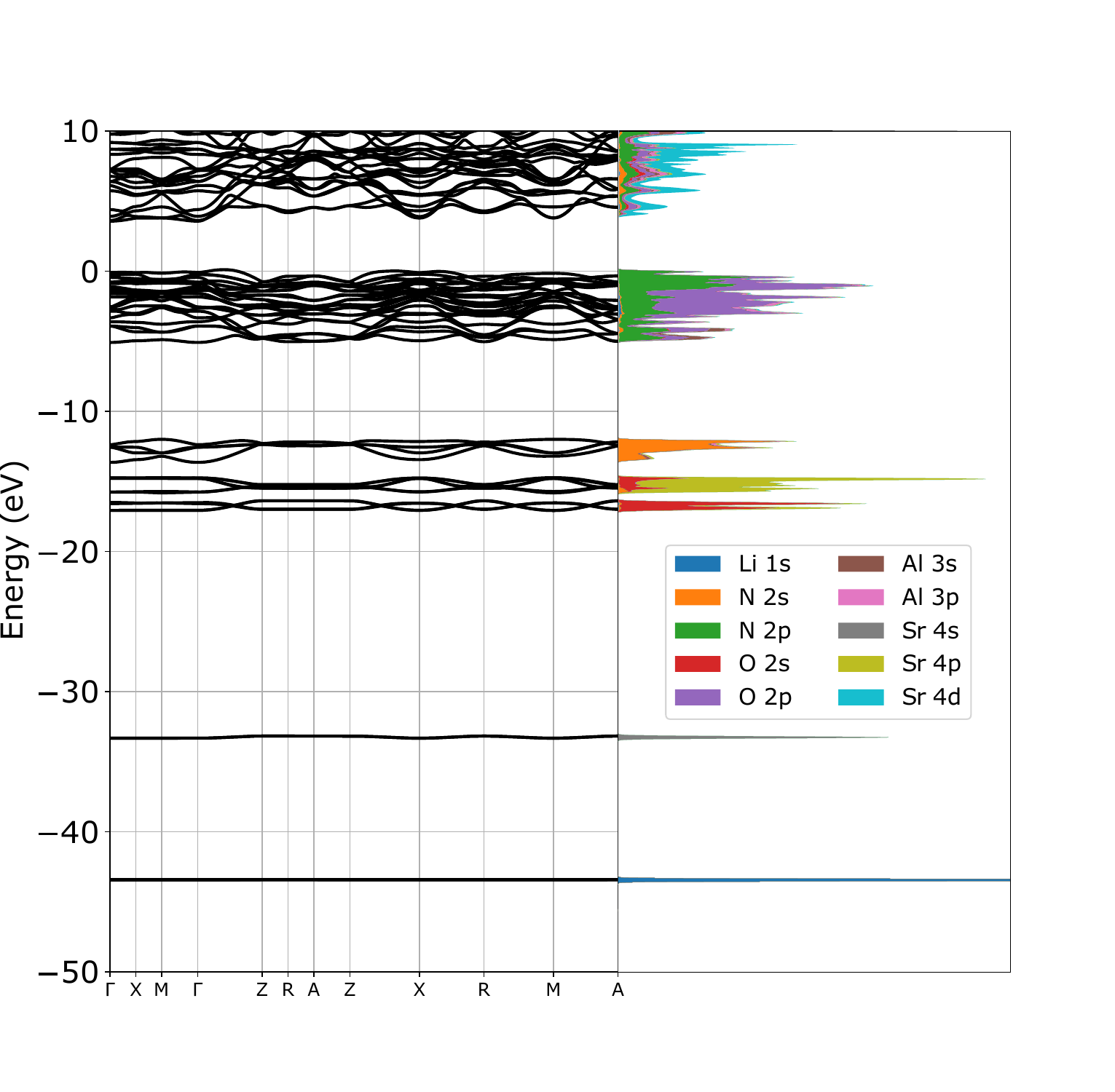}
    \caption{Left: Electronic band structure of undoped SALON in an extended energy range. Right: corresponding partial density of states. With increasing energy, we observe: deep and localized 1s states of lithium and 4s of strontium. Then three characteristic bands composed of 2s states of oxygen, 4p states of strontium, and 2s states of Nitrogen. Near the Fermi level, we have a big stack composed of 2p states of oxygen and nitrogen with 3p states of aluminium. Finally, the bottom of the conduction bands is mainly composed of 4d states of strontium.  }
    \label{fig:Complete_BS}
\end{figure}

\restoregeometry
